\documentclass[prc,showpacs,twocolumn,floatfix]{revtex4}
\usepackage{amssymb,epsfig,bm,color,slashed,mathrsfs}
\usepackage[german,english]{babel}
\newcommand{\pp}{{\rm pp}}
\newcommand{\ph}{{\rm ph}}
\newcommand{\pb}{{\rm pb}}
\newcommand{\be}{\begin{equation}}
\newcommand{\ee}{\end{equation}}
\newcommand{\bea}{\begin{eqnarray}}
\newcommand{\eea}{\end{eqnarray}}

\newcommand{\ep}{\epsilon}

\newcommand{\vecq}{{\bm q}}
\newcommand{\vecp}{{\bm p}}

\newcommand{\vecn}{{\bm n}}

\newcommand{\vecv}{\bm v}

\newcommand{\vecsig}{{\bm \sigma}}

\newcommand{\Gammavec}{{\bm \Gamma}}

\newcommand{\tauvec}{{\bm \tau}}

\newcommand{\xivec}{{\bm \xi}}
\newcommand{\ie}{{\it i.e.}}
\newcommand{\eg}{{\it e.g.}}
\def\Xint#1{\mathchoice
  {\XXint\displaystyle\textstyle{#1}}%
  {\XXint\textstyle\scriptstyle{#1}}%
  {\XXint\scriptstyle\scriptscriptstyle{#1}}%
  {\XXint\scriptscriptstyle\scriptscriptstyle{#1}}%
  \!\int}
\def\XXint#1#2#3{{\setbox0=\hbox{$#1{#2#3}{\int}$}
    \vcenter{\hbox{$#2#3$}}\kern-.5\wd0}}

\def\dashint{\Xint-}

\definecolor{red}{rgb}{0.8,0,0}
\definecolor{RED}{rgb}{0.8,0,0}
\definecolor{violet}{rgb}{0.4,0,0.4}
\definecolor{green}{rgb}{0,0.5,0.0}
\definecolor{GREEN}{rgb}{0,0.5,0.0}
\definecolor{navy}{rgb}{0.0,0.0,0.6}
\definecolor{orange}{rgb}{0.8,0.2,0.0}
\definecolor{blue}{rgb}{0.3,0.0,0.8}

\begin{document}
\title[Response functions of cold neutron matter]
{Response functions of cold neutron matter: density, spin and current fluctuations}
\author{Jochen Keller and  Armen Sedrakian} 
\affiliation{Institute for Theoretical Physics,
 J.~W.~Goethe University, D-60438  Frankfurt am Main, Germany}

\begin{abstract}
  We study the response of a single-component pair-correlated baryonic
  Fermi-liquid to density, spin, and their current perturbations.  A
  complete set of response functions is derived in the low-temperature
  regime both within an effective theory based on a small momentum
  transfer expansion and within a numerical scheme valid for arbitrary
  momentum transfers. A comparison of these two approaches validates
  the perturbative approximation within the domain of its
  convergence. We derive the spectral functions of collective
  excitations associated with the density, density-current, spin, and
  spin-current perturbations. The dispersion relations of density and
  spin fluctuations are derived and it is shown that the density
  fluctuations lead to exciton-like undamped bound states, whereas the
  spin excitations correspond to diffusive modes above the
  pair-breaking threshold. The contribution of the collective
  pair-breaking modes to the specific heat of neutron matter at
  subnuclear densities is computed and is shown to be comparable to
  that of the degenerate electron gas at not too low temperatures.
\end{abstract}

\maketitle

\section{Introduction}
\label{sec:1}

The interiors of neutron stars become superfluid shortly after their
formation (for reviews of the physics of superfluidity in neutron stars
see
Refs.~\cite{Baym:1978jf,Pethick:1995di,Lombardo:2000ec,Dean:2002zx,Sedrakian:2006xm}).
In the inner crust of a neutron star the neutrons pair in the $^1S_0$
channel with the density-dependent gap parameter in the range
$\Delta\le 1$
MeV~\cite{Lombardo:2000ec,Dean:2002zx,Sedrakian:2006xm}. The neutron
$S$-wave superfluidity persists up to the densities of order of the
nuclear saturation density $n_0$. Neutron $P$-wave superfluidity is
expected at larger densities~\cite{Pwave1,Pwave2,Pwave3}.  Protons,
which are less abundant, form an $S$-wave pair condensate from
densities $\sim n_0/2$, where they de-confine from crustal nuclei, up
to densities $n \gg n_0$, \ie, the deep interiors of the star.  For
not too large isospin asymmetries, the $D$-wave condensation of
neutron-proton pairs may set in at high densities as
well~\cite{Alm:1996zz}.


The low-energy dynamics of baryonic matter in compact stars can be
described, microscopically, in terms of a set of response functions to perturbations
having different symmetries. A frequently encountered example is the
radiation and transport of neutrinos, in which case one is interested
in vector and axial-vector perturbing operators. Response functions
also contain the complete information on the spectrum of the low-lying
excitations (\ie, density waves, spin waves, etc.) and, therefore, they
permit to evaluate the contribution of the collective excitations to
thermodynamics and transport of matter.


Near equilibrium the response functions of nuclear systems are
characterized by length scales that are large compared to the inverse
Fermi wave vector, or equivalently, energies that are small compared
to the Fermi energy.  In the unpaired limit, the Landau theory of
normal Fermi liquids provides a suitable framework for the evaluation
of response functions in compact
stars~\cite{Olsson:2002yu,Olsson:2004ea,Lykasov:2005xh,Lykasov:2008yz,Pethick:2009gj,Margueron:2005nc,Bozek:2004ct}.
The many-body problem of the evaluation of response functions entails a
number of challenges. One is the identification of the relevant set of
diagrams, when perturbation theory fails. For many systems the
response functions are computed from a resummation of an infinite number
of finite temperature ring diagrams~\cite{Negele:1988vy}. While this
scheme accounts for the (vertex renormalized) single particle-hole
excitations, it does not include multi-pair contributions to the
response functions. Such contributions are important for the
evaluation of the magnetic susceptibility of degenerate nuclear
matter~\cite{Olsson:2002yu,Olsson:2004ea,Lykasov:2005xh,Lykasov:2008yz,Pethick:2009gj}.
The second challenge is the inclusion of the non-central forces, which
arise in the nuclear systems due to the tensor forces. In fact, these
give rise to the coupling of states with more than one
quasiparticle-quasihole pair, thereby changing the static
susceptibility and the magnetic moments in the nuclear
Fermi liquid~\cite{Olsson:2002yu,Olsson:2004ea,Lykasov:2005xh,Lykasov:2008yz,Pethick:2009gj}.
As a consequence, in nuclear matter the relationship between the
Landau parameters and the magnetic susceptibility is considerably more
complicated than for systems with purely central forces. Sum-rule
arguments can be used to place a lower bound on the contribution to
the static susceptibility coming from transitions to multipair
states~\cite{Olsson:2002yu,Olsson:2004ea}. Furthermore, it was shown
that the rates of processes involving transitions to two
quasiparticle-quasihole states may be calculated in terms of the
collision integral in the Landau transport equation for
quasiparticles~\cite{Lykasov:2005xh,Lykasov:2008yz}.  The multi-loop
processes induced by the tensor forces are of paramount importance in
the astrophysics of neutron stars, since the bremsstrahlung processes
on weak neutral currents are among the leading processes contributing
to the neutrino luminosity of these
stars~\cite{Friman:1978zq,Sedrakian:2000kc,Hanhart:2000ae,Timmermans:2002hc}.

The focus of this paper is the derivation of the response functions
associated with perturbations of density, density current, spin, and
spin current in a single-component Fermi liquid. It extends our
earlier study of density response~\cite{Sedrakian:2010xe} to new types
of perturbations as well as revises some of the perturbative results
contained therein.  The energy scale characterizing the dynamical
processes in neutron stars are of the  order of temperature $T\le 1$~MeV.
 The high densities in compact stars render the Fermi energies of
fermions in the range $\ep_F \sim 10-100$ MeV. Consequently,
one needs the response functions in the limit $T/\ep_F\ll 1$. As
mentioned above the pairing gaps could be of the order of 1 MeV, \ie,
they substantially influence the dynamics of the systems for
temperatures $T\le T_c$, the critical temperature of the superfluid
phase transition.

This study is based on the method of the Green's functions for
superfluid systems at non-zero temperatures and aims at the resummation
of an infinite series of particle-hole ladder diagrams in neutron
matter. This re-summation scheme respects the gauge invariance, sum
rules, and baryon number conservation.  The appropriate technique was
first developed by Abrikosov and Gor'kov in the electrodynamics of
superconductors~\cite{AG} (see also Ref.~\cite{Abrikosov:1962}).  In this
theory the response of the superconductors to external probes is expressed
in the language of propagators at non-zero temperature and density
with contact interactions that do not distinguish among the
particle-hole and particle-particle channels. It is equivalent to the
theories initially advanced by Bogolyubov~\cite{Bogolyubov},
Anderson~\cite{Anderson} and others, which are based on the equations
of motion for second-quantized operators. Subsequently, Larkin,
Migdal, and Leggett ~\cite{Migdal:1967,Leggett:1966zz} generalized the
Landau Fermi-liquid theory to superconductors and superfluids, thus
extending the Abrikosov-Gor'kov approach to strongly interacting
regime. This last method implements the wave-function renormalization of
the quasiparticle spectrum, higher order harmonics in the interaction
channels, and postulates particle-hole ($\ph$) and particle-particle
($\pp$) interactions with different strength and/or sign.


The response functions of baryonic matter were studied in the unpaired,
but degenerate regime in the context of neutrino emission from compact
stars (see, e.g., Ref.~\cite{Sedrakian:2006mq} and references
therein).  The work on these functions in the same context, but for
superfluid baryonic matter started more
recently~\cite{Kundu:2004mz,Leinson:2006gf,Sedrakian:2006ys,Kolomeitsev:2008mc,Steiner:2008qz,Leinson:2009mq,Kolomeitsev:2010hr,Sedrakian:2012ha}.

Quite generally, the response functions to
density, spin and their current perturbations can be related to the
appropriate response functions of baryons to the operators of the
electroweak theory. 
To see the mapping explicitly consider the weak interaction
Lagrangian, which at low energies is given by \be {\cal L}_W =
-\frac{G_F}{2\sqrt{2}} (J_V^{\mu}- J_A^{\mu}) J^L_{\mu}, \ee where
$G_F$ is the Fermi constant and the vector and axial-vector currents
are defined as \bea
\label{eq:v_current}
J_V^\mu &=& c_V\bar{\Psi}_N\gamma^\mu\Psi_N 
\simeq c_V \psi_N^\dagger\left(1,\vecv_F\right)\psi_N,\\
\label{eq:a_current}
J_A^\mu &=& c_A\bar{\Psi}_N\gamma^\mu\gamma_5\Psi_N
\simeq c_A \psi_N^\dagger\left(\bm{\sigma}\vecv_F,\bm{\sigma}\right)\psi_N, 
\eea 
and
$J^L_{\mu} = \bar{\psi}\gamma_\mu(1-\gamma_5)\psi$ is the lepton
current. Here $\vecv_F$ is the Fermi velocity of baryons,
${\bm\sigma}$ is the vector of Pauli-matrices. Here and below 
the Greek indices run over  0, 1, 2, 3, and
label the temporal and three spatial coordinates; 
the spatial coordinates are also labeled by Latin indices 
and run through 1, 2, 3. 
Equations (\ref{eq:v_current}) and (\ref{eq:a_current}) approximate
the baryonic vector and axial-vector weak currents by their dominant
contributions in the non-relativistic limit by keeping the large
components of the baryonic Dirac spinors.  The bare vertices of
interest are thus given by the expression in-between the baryon fields
$\psi_N^\dagger$ and $\psi_N$ in Eqs.~(\ref{eq:v_current}) and
(\ref{eq:a_current}): \bea
\label{eq:vertices2}
\Gamma_0^{D\mu}&=&
(\Gamma_0^D,\Gammavec^D_0 )=\left(1,\vecv_F\right), \\
\Gamma_0^{S\mu}&=&
(\Gamma_0^S,\Gammavec^S_0 )=\left(\vecsig\vecv_F,\vecsig\right).  
\eea
It is now clear that there is a one-to-one correspondence between weak
interaction vertices in the non-relativistic limit and vertices
associated with the density and density-current (index $D$), as
well as the spin-current and spin-density perturbations
(index $S$).

One complication that is always present in compact stars is the fact
that the matter is multi-component in the crusts and the core of the
star.  A superfluid features Goldstone bosons associated with the
breaking of the baryon $U(1)$ number in a
superfluid~\cite{Pethick:2010zf,Cirigliano:2011tj}.  Furthermore, the
existence of the lattice of nuclei (and non-spherical nuclear phases) in
the crust adds the lattice phonons to the set of the collective modes
that propagate in the star's
crust~\cite{Sedrakian:1996,DiGallo:2011cr}.  The various modes are
coupled~\cite{Cirigliano:2011tj,Carter:2004zr,Carter:2006nd,Carter:2006mv}.
In the cores of neutron stars there are at least three fluids  $-$ the
neutron and proton Fermi liquids, which are both expected to be in the
superfluid state, and an ultra-relativistic gas of
electrons~\cite{Sedrakian:2006mq}.  The density modes associated with
the superconducting proton component in the homogeneous matter of the
outer core of neutron stars were computed in
Refs.~\cite{Baldo:2011iz,Baldo:2011nc,Baldo:2008pb}. It is clear that
our treatment of a single-component superfluid nuclear Fermi liquid
does not account for coupling among various components.  A more
complete treatment must take into account the multi-component nature
of matter.

This paper is organized as follows. The remainder of the Introduction
provides prerequisite information.  In Sec.~\ref{sec:2} the baryon
propagators and self-energies are introduced within a
finite-temperature imaginary-time theory. Vertex functions
corresponding to density and spin perturbations are discussed in
Sec.~\ref{sec:3}. Section~\ref{sec:4} is devoted to the density and
spin response functions, with two subsections discussing perturbative
expansions of response functions as well as their exact numerical
evaluation. In Sec.~\ref{sec:5} the spectral functions and collective
density and spin excitations are discussed. We evaluate the specific
heat contribution arising from these excitations in Sec.~\ref{sec:6}.
Our conclusions are collected in Sec.~\ref{sec:7}. The details of
computations are relegated to Appendices~\ref{AppendixA} and
\ref{AppendixB} and a comparison to other methods is presented in
Appendix~\ref{AppendixC}. We use the natural units $\hbar = c = 1$ and
assume that the Boltzmann constant $k_B=1$, with the exception of
Sec.~\ref{sec:6}.

\subsection{Prerequisites}

In this study we explore the temperature domain well below the
critical temperature of superfluid phase transitions; typically
$T/T_c\le 0.5$, where $T_c$ is the critical temperature of a superfluid
transition. This is the case in the dominant majority of observable
neutron stars. We further consider densities where the $^1S_0$-wave
pairing is dominant among neutrons and protons. This assumption
confines our study to the densities at and below the nuclear
saturation density. In the presumed temperature and density domain it
is safe to treat the nucleons as non-relativistic particles, \ie, the
Fermi velocity of the particles is small compared to the velocity of
light in a vacuum, $v_F\ll 1$ in natural units. This enables us to use
the non-relativistic dispersion law for the particles in the normal
state and non-relativistic limits of the Dirac matrices appearing in
the bare vertices. Furthermore, because we work in the extreme
low-temperature limit, we shall restrict the length of the momenta of
the particles to their Fermi wave-vector, \ie, we write $\vecp = m^*
v_F\vecn$, where $m^*$ is the effective mass of a quasiparticle and
$\vecn = \vecp/\vert \vecp\vert$.

One of the purposes of this work is to compare the response functions
obtained from perturbative approaches and direct numerical
computation. The perturbative treatment is based on a low-momentum
transfer expansion, where the expansion parameter is either generic
and reflects the characteristic properties of the system or is
dictated by certain kinematical conditions valid in the domain of
interest.  Examples of small parameters are $q/k_F$ or $q v_F/\omega$,
where $k_F$ is the Fermi
wave vector,  and $\omega$ and $q$ are the energy and
the magnitude of the 
momentum transfer. While the first parameter is generic for thermal
processes (\ie,~  processes in which the energy-momentum transfer is of the
order of temperature) the second is small only in the kinematical
domain of time-like processes (\eg,~ neutrino radiation).  In the
second case the momentum transfer is thermal, therefore $q \ll k_F$,
which establishes one suitable expansion parameter. We note that for
on-shell perturbations with linear spectrum, as, for example, neutrinos
($\omega = q$ in natural units) the smallness of the two expansion
parameters reduces to the condition $v_F\ll 1$, which is the same as
the non-relativistic expansion. In the case of the numerical computation
there are in principle no constraints on the values of the momentum
transfer and the Fermi wave vector.  However, since our intention is to
compare the perturbative and exact numerical results,  we will restrict
ourselves to the range of values of the parameters defined by the
perturbative treatment.

\subsection{Unpaired and pair-correlated particle spectra }
As we work in the non-relativistic limit the spectrum in the normal state is given by
\be
\xi_p=\frac{{p}^2}{2m^*}-\mu,
\ee
where $\mu$ is the chemical potential. The spectrum in the 
pair-correlated case is 
\be
\epsilon_p=\sqrt{\xi_p^2+\Delta^2},
\ee
where we assume that the gap function is momentum independent, which
is the case for contact pairing interactions. We will need frequently
the perturbed spectra of particles, which are defined in the unpaired 
case as
\bea
\xi_\pm =\frac{1}{2m^*}\left(\vecp\pm\frac{\vecq^2}{2}\right)-\mu
       \simeq \xi_p\pm\frac{\vecq\vecv}{2},
\eea
where in the second expression the small recoil term ${\vecq^2}/{8m^*}$ has
been dropped. In the paired case the quasiparticle spectrum is 
\bea 
\epsilon_\pm=\sqrt{\xi_\pm^2+\Delta^2}  \simeq \sqrt{\epsilon_p^2\pm\xi_p\vecq\vecv},
\eea
to leading order in $\vert \vecq\vert$.

\section{Baryon propagators and self-energies}
\label{sec:2}
In a normal Fermi liquid the propagator is defined as 
\be
\hat{G}_{N,\sigma\sigma'}(\bm{p},\tau-\tau')
=-\delta_{\sigma\sigma'}
 \langle T_\tau\psi_{p\sigma}(\tau)\psi^\dagger_{p\sigma'}(\tau')\rangle,
\ee
where $\tau$ is the imaginary time, $\sigma$ is the spin projection,
and $T_{\tau}$
is the time-ordering operator. The  Dyson equation for the normal
propagator is given by
\be
\hat{G}_N=\hat{G}_0+\hat{G}_0\hat{\Sigma}\hat{G}_N, 
\ee
where the index 0 refers to the free-particle propagator and
$\hat{\Sigma}$ is the self-energy.  A superfluid is described by the
following propagators
\bea
\hat{G}_{\sigma\sigma'}(\bm{p},\tau-\tau')
  &=&-\delta_{\sigma\sigma'}
     \langle T_\tau\psi_{p\sigma}(\tau)\psi^\dagger_{p\sigma'}(\tau')\rangle,\\
\hat{F} _{\sigma\sigma'}(\bm{p},\tau-\tau')
  &=&\langle T_\tau\psi_{-p\downarrow}(\tau)\psi_{p\uparrow}(\tau')\rangle,\\
\hat{F}^{+}_{\sigma\sigma'}(\bm{p},\tau-\tau')
  &=&\langle T_\tau\psi^\dagger_{p\uparrow}(\tau)\psi^\dagger_{-p\downarrow}(\tau')\rangle,\\
\hat{G}^-_{\sigma\sigma'}(\bm{p},\tau-\tau')
  &=&-\delta_{\sigma\sigma'}
     \langle T_\tau\psi^\dagger_{-p\sigma}(\tau)\psi_{-p\sigma'}(\tau')\rangle.\nonumber\\
\eea
These propagators obey Nambu-Gorkov equations 
and are given by
\bea
\hat{G} &=& \hat{G}_0+\hat{G}_0\hat{\Sigma}\hat{G}+\hat{G}_0\hat{\Delta} \hat{F}^{+}
         =\hat{G}_N+\hat{G}_N\hat{\Delta} \hat{F}^{+},\\
\hat{F}^+ &=& \hat{G}_0^-\hat{\Sigma}^-\hat{F}^{+}+\hat{G}_0^-\hat{\Delta}^{+}\hat{G}
           =\hat{G}_N^-\hat{\Delta}^{+}\hat{G},\\
\hat{F} &=& \hat{G}_0\hat{\Sigma}\hat{F} +\hat{G}_0\hat{\Delta} \hat{G}^-
         =\hat{G}_N\hat{\Delta} \hat{G}^-,\\
\hat{G}^- &=& \hat{G}_0^-+\hat{G}_0^-\hat{\Sigma}^-\hat{G}^-+\hat{G}_0^-\hat{\Delta}^+ \hat{F}
           = \hat{G}_N^-+\hat{G}_N^-\hat{\Delta}^+ \hat{F},\nonumber\\
\eea
where $\hat{G}(p)$ and $\hat{G}_0(p)$ are the full and free normal
propagators, $\hat{F} (p)$ and $\hat{F}^{+}(p)$ are the
anomalous propagators, $\hat{\Sigma}(p)$ and $\hat{\Sigma}^-(p)$ are
the normal self-energies for particles and holes, and $\hat{\Delta} (p)$ and 
$\hat{\Delta}^{+}(p)$ are the anomalous self-energies.  The
propagators and self-energies are $2\times2$-matrices in the spin
space. (We suppress the isospin space variables as we consider only
single-component ensembles with fixed isospin.)  The normal (particle
and hole) propagators and self-energies are diagonal in spin space,
\bea
 \hat{G}(p) &=& G(p)\hat{1}_2 = G^-(-p)\hat{1}_2 = \hat{G}^-(-p),\\
 \hat{\Sigma}(p) &=& \Sigma(p)\hat{1}_2 = \Sigma^-(-p)\hat{1}_2 = \hat{\Sigma}^-(-p),
\eea
while the anomalous ones are antisymmetric in spin space and therefore are
proportional to $i\sigma_2$,
\bea
 \hat{F}(p) &=& F(p)i\sigma_2, \quad   \hat{F}^{+}(p) 
             = F^{+}(p)i\sigma_2 \\
 \hat{\Delta}(p) &=& \Delta(p)i\sigma_2, \quad 
 \Delta^+(p)i\sigma_2 = \Delta^+(p)i\sigma_2,
\eea
where $\sigma_2$ stands for the second Pauli matrix. For real pairing 
gaps $\Delta(p) = \Delta^+(p)$ and $F(p) = F^+(p)$.

The propagators can be written as the sum of  a pole and a regular part
by expanding the self-energy in the vicinity of the Fermi surface.
Neglecting the (small) off-shell contributions, 
 we shall keep the pole part of the
propagators and set the wave function renormalization
$Z(p)^{-1} = 1-\partial_{\omega}\Sigma(\omega)\vert_{\omega = \xi_p} = 1$. 
The real-time solution of the Nambu-Gorkov equations 
in momentum space are
\bea
\label{Gprop1}
G &=& \frac{p_0+\xi_p}{p_0^2-\epsilon_p^2+i\eta}\nonumber\\
 & =& \frac{u_p^2}{p_0-\epsilon_p+i\eta}
  + \frac{v_p^2}{p_0+\epsilon_p+i\eta},\\
\label{Fprop1}
F &=& \frac{-\Delta}{p_0^2-\epsilon_p^2+i\eta}\nonumber\\
  &=& -u_p v_p\left(\frac{1}{p_0-\epsilon_p+i\eta}
  - \frac{1}{p_0+\epsilon_p+i\eta}\right),
\eea 
with
\bea
G(p) &=& G^-(-p),\\
\Sigma(p) &=& \Sigma^-(-p),\\
F(p) &=& F^{+}(p) = F(p)  ,\\
\Delta(p) &=& \Delta^{+}(p) = \Delta,
\eea
and the Bogolyubov amplitudes defined as 
\bea
u_p &=& \frac{1}{\sqrt{2}}\left(1+\frac{\xi_p}{\epsilon_p}\right),\\
v_p &=& \frac{1}{\sqrt{2}}\left(1-\frac{\xi_p}{\epsilon_p}\right).
\eea
The finite temperature Matsubara Green's functions are obtained via a 
replacement of the time-component of the four-momentum  in Eqs.~(\ref{Gprop1}) and
(\ref{Fprop1}) by a complex frequency 
\bea
G(ip_n,\vecp) &=& \frac{u_p^2}{ip_n-\epsilon_p} + \frac{v_p^2}{ip_n+\epsilon_p},\\
F(ip_n,\vecp) &=& -u_p v_p\left(\frac{1}{ip_n-\epsilon_p}-\frac{1}{ip_n+\epsilon_p}\right),
\eea
which assumes discrete values $p_n = (2n+1)\pi T$, where $n$ is an integer.

\section{Vertex functions}
\label{sec:3}

The equations for vertex functions involve loops which are constructed
from the convolutions of a product of two propagators.  One possible
kinematics for such products is the symmetrical one, which assigns to
an arbitrary imaginary-time propagator $X$ the arguments $X_+ = \left(
  ip_n+i\omega_m,\vecp+\frac{\vecq}{2}\right)$ and $X_- = \left(
  ip_n,\vecp-\frac{\vecq}{2}\right)$, \ie, the external momentum is
split symmetrically among the particle and the hole (but the energy
transfer is not). The remainder of this work will use this kinematics.
We now turn to the calculation of the effective (or dressed) vertices,
which take into account the modifications due to the strong
interactions in the medium. The driving interaction in the
particle-particle and particle-hole channel will be parametrized
as~\cite{Migdal:1967}
\bea
\hat V^{\pp}_{\alpha\beta\gamma\delta}
&\simeq & V^D_{\pp}(i\sigma_2)_{\alpha\beta}(i\sigma_2)_{\gamma\delta}
    +
    V^S_{\pp}({i\sigma_2\bm{\sigma}})_{\alpha\beta}\cdot
({\bm{\sigma}}i\sigma_2)_{\gamma\delta},\nonumber\\
    \\
\hat V^{\ph}_{\alpha\beta\gamma\delta}
&\simeq &V^D_{\ph}\delta_{\alpha\beta}\delta_{\gamma\delta}
    + V^S_{\ph}{\bm{\sigma}}_{\alpha\beta}\cdot{\bm{\sigma}}_{\gamma\delta}, 
\eea
where $V^D$ and $V^S$ are the interaction strengths in the density
and spin channels, the subscripts or superscripts $\pp$ and $\ph$ refer to the
particle-particle and particle-hole channels, respectively.

Since the particle momenta are restricted to the Fermi surfaces, the
amplitudes will depend only on the angle formed by the momenta of the
particles. Therefore, as in the ordinary Fermi-liquid theory, they can
be expanded in spherical harmonics with respect to this angle.  The
coefficients in this expansion are the Landau parameters. We will
retain the leading-order Landau parameter only, since the higher-order
Landau parameters are numerically insignificant. We will use below
their values for bulk neutron matter as computed 
in Ref.~\cite{Sedrakian:2006ys}.

In analogy with the random phase approximation for unpaired ensembles
the calculation of full vertices requires a summation of an infinitely
long chain of irreducible particle-hole ring diagrams. One possible
way to derive these equations is to compute the variations of the
Nambu-Gor'kov equations in an external field~\cite{Migdal:1967}.
Another method to set up the integral equations for the vertices is to
construct them directly from Feynman diagrammatic rules. 
 In any case, since a single-component superfluid
ensemble is fully described by four different propagators, one finds
that there are four topologically different vertices, which are
determined by four coupled integral equations.  
The analytical form of
these equations for scalar vertices is 
\bea 
\label{G1}
\hat{\Gamma}_1^{D/S} - \hat{\Gamma}_0^{D/S}
&=&\int\!\!\frac{d^4p}{(2\pi)^4i}\,\hat V_{\ph}^{D/S}
  \Bigl(\hat{G}\hat{\Gamma}_1^{D/S}\hat{G} \nonumber\\
&&\hspace{-1.5cm}
 + \hat{F}\hat{\Gamma}_3^{D/S}\hat{G}
  +\hat{G}\hat{\Gamma}_2^{D/S}\hat{F}
  +\hat{F}\hat{\Gamma}_4^{D/S}\hat{F}\Bigr),\\
\label{G2}
\hat{\Gamma}_2^{D/S} 
&=&\int\!\!\frac{d^4p}{(2\pi)^4i}\,\hat V_{\pp}^{D/S}
  \Bigl(\hat{G}\hat{\Gamma}_2^{D/S}\hat{G}^-\nonumber\\
&& \hspace{-1.5cm}
 + \hat{F}\hat{\Gamma}_4^{D/S}\hat{G}^-
  +\hat{G}\hat{\Gamma}_1^{D/S}\hat{F}
  +\hat{F}\hat{\Gamma}_3^{D/S}\hat{F}\Bigr),\\
\label{G3}
\hat{\Gamma}_3^{D/S} 
&=&\int\!\!\frac{d^4p}{(2\pi)^4i}\,\hat V_{\pp}^{D/S}
  \Bigl(\hat{G}^-\hat{\Gamma}_3\hat{G} \nonumber\\
&&\hspace{-1.5cm}
 + \hat{F}\hat{\Gamma}_1^{D/S}\hat{G}
  +\hat{G}^-\hat{\Gamma}_4^{D/S}\hat{F}
  +\hat{F}\hat{\Gamma}_2^{D/S}\hat{F}\Bigr),\\
\label{G4}
\hat{\Gamma}_4^{D/S} -\hat{\Gamma}_0^{D/S\,-}
&=&\int\!\!\frac{d^4p}{(2\pi)^4i}\,\hat V_{\ph}^{D/S}
\Bigl(\hat{G}^-\hat{\Gamma}_4^{D/S}\hat{G}^- \nonumber\\
&&\hspace{-1.9cm} + \hat{F}\hat{\Gamma}_1^{D/S}\hat{F}
+\hat{F}\hat{\Gamma}_2^{D/S}\hat{G}^-
+\hat{G}^-\hat{\Gamma}_3^{D/S}\hat{F}\Bigr), \eea where subscripts $D$
and $S$ refer to the density and spin, $\hat{\Gamma}_0^{D/S\,-}$ is
the bare vertex for holes.  Identical equations can be written for
vector vertices. In the following we approximate the particle-hole and
particle-particle interaction amplitudes by the leading-order Landau
parameters $v_{\ph}$ and $v_{\pp}$. The last of these is determined by the gap equation as
follows
\be 
1 = \nu v_{\pp} \int_{0}^{\Lambda}d\xi_p \frac{1-2f(\ep_p)}{2\ep_p},
\ee
where $\nu={m^*k_F}/{2\pi^2}$ is the density of states on the Fermi
surface and $\Lambda$ is the cut-off which regularizes the ultraviolet
divergence of the integral.

The solutions of the vertex equations (\ref{G1}) to (\ref{G4})
are described in Appendix~\ref{AppendixA}.  We find for 
 bare scalar vertex $\Gamma_0^D=1$
\bea\label{eq:FGamma1}
\Gamma_1^D(\omega,\vecq)
=\Gamma_4^D(\omega,\vecq)
 = \frac{\mathcal{C}(\omega,\vecq)}
        {\mathcal{C}(\omega,\vecq) - v_{\ph}^D \mathcal{Q}^+(\omega,\vecq)},\\
\label{eq:FGamma2}
\Gamma_3^D(\omega,\vecq)
=-\Gamma^D_2(\omega,\vecq)
 = \frac{\mathcal{D}^+(\omega,\vecq)}
        {\mathcal{C}(\omega,\vecq) - v_{\ph}^D
          \mathcal{Q}^+(\omega,\vecq)}, 
\eea
for the bare vector vertex ${\bm{\Gamma}}_0^D = \vecv$ 
\bea
\label{eq:FGamma3}
{\bm{\Gamma}}_{1/4}^D(\omega,\vecq)
&=&\left[\pm\vecn_v
    +\frac{v_{\ph}^D \tilde{\mathcal{Q}}^-(\omega,\vecq)}        
          {\mathcal{C}(\omega,\vecq) - v_{\ph}^D \mathcal{Q}^+(\omega,\vecq)}
    \vecn_\vecq\right]\,v_F,\nonumber\\\\
\label{eq:FGamma4}
{\bm{\Gamma}}_{2/3}^D(\omega,\vecq)
&=&
\pm\Bigg[\tilde{\mathcal{D}}^-(\omega,\vecq)\nonumber\\
    &+&\frac{v_{\ph}^D \mathcal{D}^+(\omega,\vecq)\tilde {\mathcal{Q}}^-(\omega,\vecq)}
          {\mathcal{C}(\omega,\vecq) - v_{\ph}^D {\mathcal{Q}}^+(\omega,\vecq)}
   \Bigg]\,\frac{\vecn_\vecq v_F}{\mathcal{C}(\omega,\vecq)},\nonumber\\
\eea
for the bare scalar spin-current vertex $\Gamma_0^S={\bm{\sigma}}\vecv$  
\bea
\label{eq:FGamma5}
{\bm{\Gamma}}_{1/4}^S(\omega,\vecq)
&=&
{\bm{\sigma}}\left[\vecn_v
   \pm\frac{v_{\ph}^S\tilde{\mathcal{Q}}^+(\omega,\vecq) \vecn_\vecq}
           {\mathcal{C}(\omega,\vecq) - v_{\ph}^S
             \mathcal{Q}^-(\omega,\vecq)}
   \right]\,v_F,\nonumber\\\\
\label{eq:FGamma6}
{\bm{\Gamma}}_{2/3}^S(\omega,\vecq)
&=&\pm \Bigg[\tilde{\mathcal{D}}^+(\omega,\vecq)
\nonumber\\
   &+&\frac{v_{\ph}^S \mathcal{D}^-(\omega,\vecq) 
\tilde {\mathcal{Q}}^+(\omega,\vecq)}
         {\mathcal{C}(\omega,\vecq) - v_{\ph}^S \mathcal{Q}^-(\omega,\vecq)}
   \Bigg]\frac{{\bm{\sigma}}\vecn_\vecq\,v_F}{\mathcal{C}(\omega,\vecq)},
\eea
and, finally, for the bare spin vertex $\bm{\Gamma}_0^S={\bm{\sigma}}$ 
\bea\label{eq:FGamma7}
{\bm{\Gamma}}_1^S(\omega,\vecq)
=-{\bm{\Gamma}}_4^S(\omega,\vecq)
 = \frac{{\bm{\sigma}}\,\mathcal{C}(\omega,\vecq)}
        {\mathcal{C}(\omega,\vecq) - v_{\ph}^S \mathcal{Q}^-(\omega,\vecq)},\\
\label{eq:FGamma8}
        {\bm{\Gamma}}_3^S(\omega,\vecq)
        =-{\bm{\Gamma}}_2^S(\omega,\vecq) =
        \frac{{\bm{\sigma}}\,\mathcal{D}^-(\omega,\vecq)}
        {\mathcal{C}(\omega,\vecq) - v_{\ph}^S
          \mathcal{Q}^-(\omega,\vecq)}.  
\eea 
The functions on the right-hand side of
Eqs.~(\ref{eq:FGamma1}) to (\ref{eq:FGamma8})
are defined in  Appendix~\ref{AppendixA}.
The full vertex entering the density response is seen to coincide with
the one derived in Refs.~\cite{Sedrakian:2006ys,Sedrakian:2010xe,Kolomeitsev:2010hr}.
The remainder vertices are in agreement with the ones obtained in
Ref.~\cite{Kolomeitsev:2010hr}.

\section{Response functions}
\label{sec:4}

We start with a general expression for a response function in terms of
a current-current correlation function
\be\label{eq:res_gen1}
 \Pi^{\mu\nu} = \frac{1}{2}\int\!\frac{d^4p}{(2\pi)^4i}{\rm Tr}\,\left\{\hat{J}_0^\mu\hat{J}^\nu\right\}\,,
\ee
where $\hat{J}_0^\mu$ and $\hat{J}^\nu$ are the bare and dressed
currents. 
The polarization tensor consists of four different contributions
(we drop here the subscripts $D/S$)
\bea\label{eq:res_gen2}
\Pi^{\mu\nu}
&=&\frac{1}{2}\int\!\frac{d^4p}{(2\pi)^4i}\,
{\rm Tr}\,\Big[\hat{\Gamma}_0^\mu \hat{G}\left(p+\frac{q}{2}\right)
          \hat{\Gamma}_1^\nu \hat{G}\left(p-\frac{q}{2}\right)\Big]
\nonumber\\
&+&\frac{1}{2}\int\!\frac{d^4p}{(2\pi)^4i}\,
{\rm Tr}\,\Big[\hat{\Gamma}_0^\mu \hat{G}\left(p+\frac{q}{2}\right)
          \hat{\Gamma}_2^\nu \hat{F}\left(p-\frac{q}{2}\right)\Big]
\nonumber\\
&+&\frac{1}{2}\int\!\frac{d^4p}{(2\pi)^4i}\,
{\rm Tr}\,\Big[\hat{\Gamma}_0^\mu \hat{F}\left(p+\frac{q}{2}\right)
          \hat{\Gamma}_3^\nu \hat{G}\left(p-\frac{q}{2}\right)\Big]
\nonumber\\
&+&\frac{1}{2}\int\!\frac{d^4p}{(2\pi)^4i}\,
{\rm Tr}\,\Big[\hat{\Gamma}_0^\mu \hat{F}\left(p+\frac{q}{2}\right)
          \hat{\Gamma}_4^\nu \hat{F}\left(p-\frac{q}{2}\right)\Big].\nonumber\\
\eea
The trace should be carried out in the spin space. We can
now compute the response functions by substituting the bare and
effective vertices corresponding to the desired type of
perturbation. For the density response the vertices are 
$\hat{{{\Gamma}}}_{0D}^{0} $ and $\hat{{{\Gamma}}}_{jD}^{0}$ and we find
\bea\label{eq:res_D00}
\Pi_{D}^{00} (\omega,\vecq) 
&=&\frac{\mathcal{Q}^+(\omega,\vecq)}
        {\mathcal{C}(\omega,\vecq) - v_{\ph}^D \mathcal{Q}^+(\omega,\vecq)}.
\eea
Furthermore, the density-current response is given by (summation over repeated
indices is assumed)
\bea\label{eq:res_Djj}
\Pi_{D}^{jj}(\omega,\vecq)
&=&\Biggl\{
   \mathcal{A}^-(\omega,\vecq)
  -\frac{\tilde{\cal B}(\omega,\vecq)\tilde{\mathcal{D}}^-(\omega,\vecq)}
        {\mathcal{C}(\omega,\vecq)}
  \nonumber\\
 &&\hspace{-1cm}+
  \frac{v_{\ph}^D\tilde{\mathcal{Q}}^+(\omega,\vecq)\tilde{\mathcal{Q}}^-(\omega,\vecq)}
        {\mathcal{C}(\omega,\vecq)^2 - v_{\ph}^D \mathcal{C}(\omega,\vecq)\mathcal{Q}^+(\omega,\vecq)}
   \Biggr\}\,v_F^2 ,
\eea
the spin-current response is given by 
\bea\label{eq:res_S00}
\Pi_{S}^{00}(\omega,\vecq)
&=&\Biggl\{
  \mathcal{A}^+(\omega,\vecq)
 -\frac{\tilde{\cal B}(\omega,\vecq)\tilde{\mathcal{D}}^+(\omega,\vecq)}
       {\mathcal{C}(\omega,\vecq)}\nonumber\\
& &\hspace{-2cm}+\frac{v_{\ph}^S\tilde{\mathcal{Q}}^-(\omega,\vecq)\tilde{\mathcal{Q}}^+(\omega,\vecq)}
       {\mathcal{C}(\omega,\vecq)^2 - v_{\ph}^S \mathcal{C}(\omega,\vecq)\mathcal{Q}^-(\omega,\vecq)}
  \Biggr\}\,v_F^2, 
\eea
and finally, the spin-density response is 
\bea\label{eq:res_Sjj}
\Pi_{S}^{jj}(\omega,\vecq)
&=&\frac{3\mathcal{Q}^-(\omega,\vecq)}
       {{\cal C}(\omega,\vecq)-v_{\ph}^S\mathcal{Q}^-(\omega,\vecq)}.
\eea
The functions appearing on the right-hand side of
Eqs.~(\ref{eq:res_Djj}) to (\ref{eq:res_Sjj}) 
are defined in Appendix~\ref{AppendixA}.
For density perturbations the off-diagonal elements of the
polarization tensor with mixed temporal and spatial indices 
are given by (below for the sake of brevity we drop the arguments of
the loops)
\bea
\label{eq:res_Di0}
\Pi_{D}^{i0}(\omega,\vecq)
&=&\int\!\frac{d\Omega}{4\pi}\,
\Bigg\{
        \frac{A^{+}\mathcal{C}-B\mathcal{D}^{+}}
             {\mathcal{C}-v_{\ph}^S\,\mathcal{Q}^{+}}
         \Bigg\}\vecn_v^i \,v_F,
\eea
and
\bea
\label{eq:res_D0j}
 \Pi_{D}^{0j}(\omega,\vecq)
 &=&\int\!\frac{d\Omega}{4\pi}\,
          \Bigg\{
           A^{-}\,\vecn_v^j
          -B\,\left(\frac{\tilde{\mathcal{D}}^{-}}{\mathcal{C}}\right)\,\vecn_q^j\nonumber\\
          &+&\frac{v_{\ph}^D\Big(A^{+}\mathcal{C}-B\mathcal{D}^{+}\Big)\,\tilde{\mathcal{Q}}^{-}}
                      {\mathcal{C}^2-v_{\ph}^D\mathcal{C}\mathcal{Q}^{+}}
           \,\vecn_q^j
          \Bigg\}\,v_F\,,
\eea
while for spin-perturbations they are given by 
\bea
\label{eq:res_Si0}
\Pi_{S}^{i0}(\omega,\vecq)
&=&\int\!\frac{d\Omega}{4\pi}\,
         \Bigg\{
          A^{+}\,\vecn_v^i
         -B\,
           \left(\frac{\tilde{\mathcal{D}}^{+}}{\mathcal{C}}\right)\,\vecn_q^i\nonumber\\
        & +&\frac{v_{\ph}^S\Big(A^{-}\mathcal{C}-B\mathcal{D}^{-}\Big)\,\tilde{\mathcal{Q}}^{+}}
                     {\mathcal{C}^2-v_{\ph}^S\mathcal{C}\mathcal{Q}^{-}}
          \,\vecn_q^i
         \Bigg\}\,v_F,
\eea
\begin{equation}
\label{eq:res_S0j}
\Pi_{S}^{0j}(\omega,\vecq)
=\int\!\frac{d\Omega}{4\pi}
       \Bigg\{
        \frac{A^{-}\mathcal{C}-B\mathcal{D}^{-}}
             {\mathcal{C}-v_{\ph}^S\,\mathcal{Q}^{-}}
       \Bigg\}\,\vecn_v^j\,v_F\,.
\end{equation}
Each of the polarization tensors can be decomposed into transverse
and  longitudinal parts with respect to the direction of the momentum
transfer $\vecq$ according to
\begin{eqnarray}
\label{eq:res_L}
 \Pi_L(\omega,\vecq)
&=&\Pi^{00}(\omega,\vecq)\,,\\
 \Pi_T(\omega,\vecq)
&=&\frac{1}{2}\Big(\delta^{ij}-\vecn_q^i\vecn_q^j\Big)\,\Pi^{ij}(\omega,\vecq)\,.
\end{eqnarray}
Performing the decomposition of the vector polarization tensor we
obtain for the longitudinal projection
\begin{eqnarray}
\label{eq:VL}
  \Pi_{V,L}(\omega,\vecq)
&=&\frac{\mathcal{Q}^{+}(\omega,\vecq)}
       {\mathcal{C}(\omega,\vecq) - v_{\ph}^D \mathcal{Q}^{+}(\omega,\vecq)}
\end{eqnarray}
and for transverse projection
\bea
\label{eq:res_VT}
\Pi_{V,T}(\omega,\vecq)=\frac{v_F^2}{2}
  \int\!\frac{d\Omega}{4\pi}\,
          \Bigg\{
           A^{-}(\omega,\vecq)
          -A^{-}(\omega,\vecq)\,(\vecn_v^j\vecn_q^j)^2
          \Bigg\}.\nonumber\\
\eea
The longitudinal and transverse components of the axial-vector
polarization read
\bea
\label{eq:res_AL}
 \Pi_{A,L}(\omega,\vecq)
&=&\Biggl\{
  \mathcal{A}^{+}(\omega,\vecq)
 -\frac{\tilde{\mathcal{B}}(\omega,\vecq)\tilde{\mathcal{D}}^{+}(\omega,\vecq)}
       {\mathcal{C}(\omega,\vecq)}\nonumber\\
 &&\hspace{-1cm}+\frac{v_{\ph}^S\tilde{\mathcal{Q}}^{-}(\omega,\vecq)\tilde{\mathcal{Q}}^{+}(\omega,\vecq)}
       {\mathcal{C}(\omega,\vecq)^2 - v_{\ph}^S \mathcal{C}(\omega,\vecq)\mathcal{Q}^{-}(\omega,\vecq)}
 \Biggr\}\,v_F^2\,,\\
\label{eq:res_AT}
 \Pi_{A,T}(\omega,\vecq)
&=&\frac{\mathcal{Q}^{-}(\omega,\vecq)}
       {\mathcal{C}(\omega,\vecq)-v_{\ph}^S\,\mathcal{Q}^{-}(\omega,\vecq)}\,.
\end{eqnarray}
These results, which are valid for arbitrary orientations of the 
external vectors fields, can be further simplified by a suitable choice of 
the coordinate system.

\subsection{Perturbative results}
We now expand the loop functions with respect to the small parameter
$y=q/k_F$ and keep contributions up to fourth order in this parameter.
The thermal function ${\cal G}$ depends on $y$ and 
$x \equiv \vecn_{\vecq}\cdot \vecn_{\vecv}$, therefore, we can write 
\be
\mathcal{G} = \sum\limits_{k=0}^{\infty}
\mathcal{G}_{2k}\,x^{2k}\,y^{2k}
= \mathcal{G}_0+\mathcal{G}_2x^2y^2
+ \mathcal{G}_4\,x^4y^4+\mathcal{O}(y^6).
\ee
The expansions of the loop functions contain only even functions of
the parameter $y$, since possible odd terms will disappear after angle
integration; thus, \eg, for the $\cal A$-loop we obtain
\be
\label{eq:Aexp}
\mathcal{A} = \mathcal{A}_0+\mathcal{A}_2\,y^2+\mathcal{A}_4\,y^4+\mathcal{O}(y^6),
\ee and similarly for the other three.  In practice, we expand the
pre-factors in Eqs.~(\ref{eq:LoopA}) to (\ref{eq:LoopD}) as well as the
function ${\cal G}$ in the power series in parameter $y$ and subsequently
combine them.  This leads us to the following explicit expressions:
\bea 
\nu^{-1}\mathcal{A}^+ 
&=&-\int\!\!\frac{d\Omega}{4\pi} \left[1+\frac{4\mu^2x^2}{\omega^2}y^2
  +\frac{16\mu^4x^4}{\omega^4}y^4\right]\mathcal{G}(\vecv,\vecq\vecv,\vecq)
  \nonumber\\
 &&-\Biggl[\mathcal{G}_0 +\left(\frac{4\mu^2}{3\omega^2}\mathcal{G}_0
    +\frac{1}{3}\mathcal{G}_2\right)y^2 \nonumber\\
  &&+\left(\frac{16\mu^4}{5\omega^4}\mathcal{G}_0
    +\frac{4\mu^2}{5\omega^2}\mathcal{G}_2
    +\frac{1}{5}\mathcal{G}_4\right)y^4
\Biggr],\\
\nu^{-1}\mathcal{A}^-
&=&-\int\!\frac{d\Omega}{4\pi}\left[\frac{4\mu^2x^2}{\omega^2}y^2
  +\frac{16\mu^4x^4}{\omega^4}y^4\right]\,\mathcal{G}(\vecv,\vecq\vecv,\vecq)
     \nonumber\\
 &&-\left[\frac{4\mu^2}{3\omega^2}\mathcal{\mathcal{G}}_0y^2
         +\left(\frac{16\mu^4}{5\omega^4}\mathcal{G}_0
               +\frac{4\mu^2}{5\omega^2}\mathcal{G}_2\right)y^4\right],
  \eea
where we have dropped terms $\mathcal{O}(y^5)$ and higher.
Note that the term $\mathcal{G}(\vecq,\vecv,\vecq\vecv)$ is purely 
real, \ie , 
does not contribute to the imaginary parts of the
loops. For fixed momentum transfer it is constant and yields
numerically negligible contribution.
For the remaining loops we obtain 
\bea
\nu^{-1}\mathcal{B}
&=&-\frac{\omega}{2\Delta}\,\mathcal{\mathcal{G}}_0
  -\frac{\omega}{6\Delta}\mathcal{\mathcal{G}}_2y^2
  -\frac{\omega}{10\Delta}\mathcal{\mathcal{G}}_4y^4,\\
\nu^{-1}\mathcal{C}
&=&\frac{\omega^2}{4\Delta^2}\,\mathcal{\mathcal{G}}_0
  +\Big(-\frac{\mu^2}{3\Delta^2}\,\mathcal{G}_0
         +\frac{\omega^2}{12\Delta^2}\,\mathcal{G}_2\Big)y^2
  \nonumber\\
  &+&\Big(-\frac{\mu^2}{5\Delta^2}\,\mathcal{G}_2
         +\frac{\omega^2}{20\Delta^2}\,\mathcal{G}_4\Big)\,y^4,\\
\nu^{-1} \mathcal{D}^+
&=&\frac{\omega}{2\Delta}\,\mathcal{\mathcal{G}}_0
  +\frac{\omega}{6\Delta}\,\mathcal{\mathcal{G}}_2y^2
  +\frac{\omega}{10\Delta}\,\mathcal{\mathcal{G}}_4\,y^4,\\
\nu^{-1}\mathcal{D}^-
&=&\int\!\frac{d{{\Omega}}}{4\pi}
  \left[\frac{\mu x}{\Delta}\,\mathcal{G}_0\,y
  +\frac{\mu x^3}{\Delta}\,\mathcal{G}_3\,y^3\right]=0.
  \eea
One can now readily identify the coefficients of the expansion
(\ref{eq:Aexp}) and its counterparts for the remaining loops. 
In full analogy,  an expansion of the polarization tensors is given as
\be
\Pi_{D/S}^{\mu\nu}=\Pi_{D/S,0}^{\mu\nu}+\Pi_{D/S,2}^{\mu\nu}y^2
+\Pi_{D/S,4}^{\mu\nu}\,y^4+\mathcal{O}(y^6).
\ee
The coefficients of the density response function are
\bea
\Pi_{D0}^{00}
&=&\frac{\mathcal{Q}_0}
        {\mathcal{C}_0-v_{\ph}^D
\mathcal{Q}_0
}
,\\
\Pi_{D2}^{00}
&=&\frac{\mathcal{A}^+_2\mathcal{C}_0^2
         -\mathcal{B}_2\mathcal{C}_0\mathcal{D}^+_0
         +\mathcal{B}_0\mathcal{C}_2\mathcal{D}^+_0
         -\mathcal{B}_0\mathcal{C}_0\mathcal{D}^+_2}
        {\Big[\mathcal{C}_0-v_{\ph}^D \mathcal{Q}_0\Big]^2},\nonumber\\\\
\Pi_{D4}^{00}
&=&\frac{\mathcal{A}^+_4\mathcal{C}_0+\mathcal{A}^+_2\mathcal{C}_2+\mathcal{A}^+_0\mathcal{C}_4}
          {\mathcal{C}_0-v_{\ph}^D\mathcal{Q}_0}\nonumber\\
&-&\frac{
         \mathcal{B}_4\mathcal{D}^+_0+\mathcal{B}_2\mathcal{D}^+_2+\mathcal{B}_0\mathcal{D}^+_4}
          {\mathcal{C}_0-v_{\ph}^D\mathcal{Q}_0}\nonumber\\
  &-&\frac{(\mathcal{A}^+_2\mathcal{C}_0+\mathcal{A}^+_0\mathcal{C}_2
                 -\mathcal{B}_2\mathcal{D}^+_0-\mathcal{B}_0\mathcal{D}^+_2) \mathscr{C}_2} 
          {\Big[\mathcal{C}_0-v_{\ph}^D \mathcal{Q}_0\Big]^2}
  \nonumber\\
  &+&\frac{\big(\mathcal{A}^+_0\mathcal{C}_0-\mathcal{B}_0\mathcal{D}^+_0\big)
            \mathscr{C}_2^2}
          {\Big[\mathcal{C}_0-v_{\ph}^D\mathcal{Q}_0\Big]^3}\nonumber\\
   &-&\frac{\big(\mathcal{A}^+_0\mathcal{C}_0-\mathcal{B}_0\mathcal{D}^+_0\big)\mathscr{C}_4  }
         {\Big[\mathcal{C}_0-v_{\ph}^D\mathcal{Q}_0\Big]^2},
\eea
where
\bea 
\mathcal{Q}_0 &=& \mathcal{A}^+_0\mathcal{C}_0-\mathcal{B}_0\mathcal{D}^+_0,\\
\mathscr{C}_2 &=& 
\mathcal{C}_2-v_{\ph}
              \big(\mathcal{A}^+_2\mathcal{C}_0+\mathcal{A}^+_0\mathcal{C}_2
                  -\mathcal{B}_2\mathcal{D}^+_0-\mathcal{B}_0\mathcal{D}^+_2\big),\nonumber\\\\
\mathscr{C}_4 &=& \mathcal{C}_4-v_{\ph}
              \big(\mathcal{A}^+_4\mathcal{C}_0+\mathcal{A}^+_2\mathcal{C}_2+\mathcal{A}^+_0\mathcal{C}_4
\nonumber\\
                  &-&\mathcal{B}_4\mathcal{D}^+_0-\mathcal{B}_2\mathcal{D}^+_2-\mathcal{B}_0\mathcal{D}^+_4\big).
\eea
However, it turns out that $\mathcal{Q}_0 =0$ and 
\bea
&&\mathcal{A}^+_2\mathcal{C}_0^2-\mathcal{B}_2\mathcal{C}_0\mathcal{D}^+_0
         +\mathcal{B}_0\mathcal{C}_2\mathcal{D}^+_0-\mathcal{B}_0\mathcal{C}_0\mathcal{D}^+_2=0.
\eea
Consequently,  the expansion coefficients of the polarization tensor are 
\bea
\Pi_{D0}^{00}&=&0,\\
\Pi_{D2}^{00}&=&0,\\
\Pi_{D4}^{00}
&=&\frac{1}{\mathcal{C}_0}
   \bigg(\mathcal{A}^+_4\mathcal{C}_0+\mathcal{A}^+_2\mathcal{C}_2+\mathcal{A}^+_0\mathcal{C}_4\nonumber\\
        &-&\mathcal{B}_4\mathcal{D}^+_0-\mathcal{B}_2\mathcal{D}^+_2-\mathcal{B}_0\mathcal{D}^+_4\bigg), 
\eea
\ie, the density response function obtains a non-zero contribution at
order $y^4$. The coefficients for the current response are 
\bea
\Pi_{D0}^{jj}&=&0,\\
\Pi_{D2}^{jj}
&=&\Bigg\{\mathcal{A}^-_2
  -\frac{\tilde{\mathcal{B}}_1\tilde{\mathcal{D}}^-_1}{\mathcal{C}_0}
  +\frac{v_{\ph}^D\,\tilde{\mathcal{A}}^-_1\mathcal{C}_0
         \big(\tilde{\mathcal{A}}^+_1\mathcal{C}_0
              -\tilde{\mathcal{B}}_1\mathcal{D}^+_0\big)}
        {\mathcal{C}_0}\Bigg\}
   \,v_F^2,\nonumber\\\\
\Pi_{D4}^{jj}
   &=&\Bigg\{\mathcal{A}^-_4
     -\frac{\tilde{\mathcal{B}}_1\mathcal{C}_0\tilde{\mathcal{D}}^-_3
           -\tilde{\mathcal{B}}_1\mathcal{C}_2\tilde{\mathcal{D}}^-_1
           +\tilde{\mathcal{B}}_3\mathcal{C}_0\tilde{\mathcal{D}}^-_1}
           {\mathcal{C}_0^2}\nonumber\\
   &+&v_{\ph}^D\Bigg[
       \frac{\big(\tilde{\mathcal{A}}^-_3\mathcal{C}_0^2
                 +2\tilde{\mathcal{A}}^-_1\mathcal{C}_0\mathcal{C}_2
                 -\tilde{\mathcal{A}}^-_1 \mathscr{C}\big)
             \big(\tilde{\mathcal{A}}^+_1\mathcal{C}_0
                 -\tilde{\mathcal{B}}_1\mathcal{D}^+_0\big)}
           {\mathcal{C}_0^2}
   \nonumber\\
   &&+\tilde{\mathcal{A}}^-_1\,
             \Big(\tilde{\mathcal{A}}^+_1\mathcal{C}_2
                  +\tilde{\mathcal{A}}^+_3\mathcal{C}_0
                  -\tilde{\mathcal{B}}_1\mathcal{D}^+_2
                  -\tilde{\mathcal{B}}_3\mathcal{D}^+_0\Big)
      \Bigg]\Bigg\}
   \,v_F^2,\nonumber\\
\eea
where 
\bea
\mathscr{C} &=& 2\mathcal{C}_0\mathcal{C}_2-v_{\ph}^D
                  \Big(\mathcal{A}^+_2\mathcal{C}_0^2
                        +\mathcal{A}^+_0\mathcal{C}_0\mathcal{C}_2
                        -\mathcal{B}_0\mathcal{C}_0\mathcal{D}^+_2\nonumber\\
                        &-&\mathcal{B}_0\mathcal{C}_2\mathcal{D}^+_0
                        -\mathcal{B}_2\mathcal{C}_0\mathcal{D}^+_0\Big).
\eea
In the case of the current response the first non-zero term arises at
the order $y^2$ and the fourth order term is  sub-leading.

Note that the vector current polarization tensor must vanish at the zeroth order 
as required by the  $f$-sum rule~\cite{Negele:1988vy}
\be
\lim_{\vecq\to 0}\int\!d\omega\,\omega\,{{\rm Im}}\Pi^D(\vecq,\omega) = 0.
\ee
This is a direct consequence of the conservation of the baryon number.
For the spin-current response we find
\bea\label{eq:PiSC1}
\Pi_{S0}^{00}&=&\mathcal{A}^+_0\,v_F^2,\\
\label{eq:PiSC2}
\Pi_{S2}^{00}&=&\bigg\{\mathcal{A}^+_2
                +v_{\ph}^S
                \Big(\tilde{\mathcal{A}}^+_1\tilde{\mathcal{A}}^-_1\mathcal{C}_0
                     \tilde{\mathcal{A}}^+_1\tilde{\mathcal{A}}^-_1\mathcal{C}_0\Big)
                \bigg\}\,v_F^2,\\
\label{eq:PiSC3}
\Pi_{S4}^{00}&=&\bigg\{\mathcal{A}^+_4
                +v_{\ph}^S
                \Big(\tilde{\mathcal{A}}^+_3\tilde{\mathcal{A}}^-_1\mathcal{C}_0
                    +\tilde{\mathcal{A}}^+_1\tilde{\mathcal{A}}^-_3\mathcal{C}_0
                    +\tilde{\mathcal{A}}^+_1\tilde{\mathcal{A}}^-_1\mathcal{C}_2
                \nonumber\\
              &&
                    -\tilde{\mathcal{A}}^-_3\tilde{\mathcal{B}}_1\mathcal{D}^{+}_0
                    -\tilde{\mathcal{A}}^-_1\tilde{\mathcal{B}}_3\mathcal{D}^{+}_0
                    -\tilde{\mathcal{A}}^-_1\tilde{\mathcal{B}}_1\mathcal{D}^{+}_2
                \Big)\nonumber\\
             &&
                +(v_{\ph}^S)^2
                \Big(\tilde{\mathcal{A}}^+_1\tilde{\mathcal{A}}^-_1\mathcal{A}^-_{2}\mathcal{C}_0               
                    +\tilde{\mathcal{A}}^+_1\tilde{\mathcal{B}}_1\mathcal{A}^-_{2}\mathcal{C}_0\Big)
                \bigg\}\,v_F^2.               \nonumber\\
\eea
In this case the leading order contribution, given by Eq.~(\ref{eq:PiSC1}),
is of order $y^0$ in the $y$-expansion, which means that the vertex
corrections introduce sub-leading order corrections and the
single-loop result is a good approximation to the full polarization
tensor.  Finally, the spin-density response is given by
\bea
\Pi_{S0}^{jj}&=&0,\\
\Pi_{S2}^{jj}&=&3\,\mathcal{A}^-_2\,v_F^2,\\
\Pi_{S4}^{jj}&=&3\left(\mathcal{A}^-_4+v_{\ph}^S\mathcal{A}^{-\,2}_2\right)\,v_F^2. 
\eea
The leading order contribution now arises at order $y^2$.

If we restrict ourselves only to the leading order contributions in
each channel, then these contain only the leading order term
in the expansion of the thermal function ${\cal G}_0$, \ie , at order
$y^0$.
The explicit expressions are 
\bea
\label{eq:pi_density}
\Pi_D^{00}
&=&-\frac{64\mu^4}{45\omega^4}\,y^4\,\mathcal{G}_0
 = -\frac{4 q^4 v_F^4}{45\omega^4}\,\mathcal{G}_0,\\
\label{eq:pi_current}
\Pi_D^{jj}
&=&-\frac{8\mu^2 v_F^2}{9\omega^2}\,y^2\,\mathcal{G}_0
 = -\frac{2 q^2 v_F^4}{9\omega^2}\,\mathcal{G}_0,\\
\label{eq:pi_spincurrent}
\Pi_S^{00}&=&-v_F^2\,\mathcal{G}_0,\\
\label{eq:pi_spin}
\Pi_S^{jj}
&=&-\frac{4\mu^2}{\omega^2}\,y^2\,\mathcal{G}_0\,
 = -\frac{q^2 v_F^2}{\omega^2}\,\mathcal{G}_0 .
\eea
The last equalities in these expressions make it clear that the
expansion, which was initially carried out with respect to the
parameter $q/k_F$ maps onto the expansion in $v_F$. It is seen that the
vector current polarization tensors are of order $O(v_F^4)$ while the
axial vector polarization tensors are of order $O(v_F^2)$. The
perturbative results (\ref{eq:pi_density}) and (\ref{eq:pi_current})
are in good agreement with the ones derived recently in the context of
vector neutrino
emission~\cite{Leinson:2006gf,Kolomeitsev:2010hr,Sedrakian:2012ha}. Similarly,
the perturbative expressions in the spin channel 
(\ref{eq:pi_spincurrent}) and (\ref{eq:pi_spin}) are in agreement with
the original results derived in the context of the axial vector
neutrino 
emission~\cite{Flowers:1976ux,Yakovlev:1998wr,Kaminker:1999ez,Kolomeitsev:2010hr}.

\subsection{Numerical results for response functions}

\begin{figure}[t]
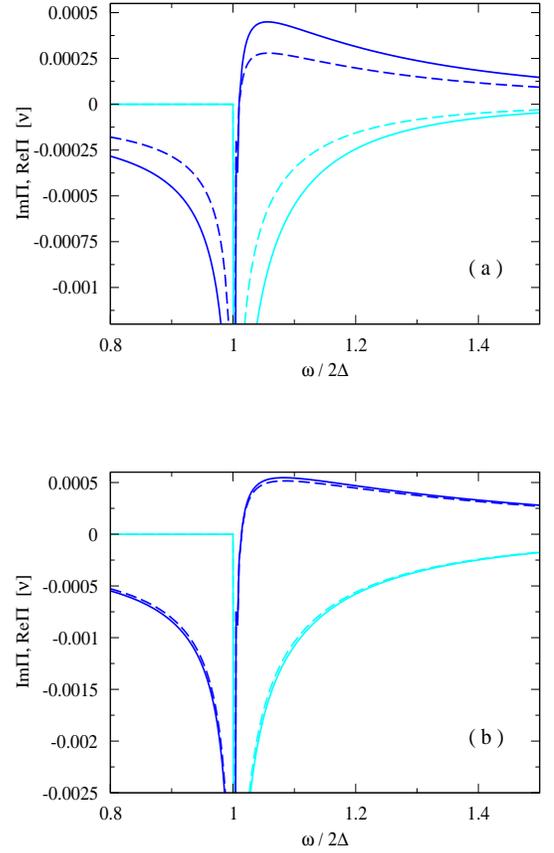

\begin{center}
\includegraphics[width=7.0cm,height=5.0cm,angle=0]{density_kf10_05tc.eps}
\vskip 1.2cm
\includegraphics[width=7.0cm,height=5.0cm,angle=0]{current_kf10_05tc.eps}
\caption[] {(Color online) Numerical (solid lines) and perturbative
  (dashed lines) results for the imaginary (heavy, blue line) and the
  real (light, cyan line) parts of the (a) density response function
  $\Pi^D_{00}$ and (b) the density-current response function
  $\Pi^D_{jj}$ normalized to the density of states $\nu$ in units
  fm$^{-2}$. The energy transfer $\omega$ is in units of $2\Delta(T)$.
  The temperature is fixed at $0.5\,T_c$ with pairing gap
  $\Delta=1.0~\textrm{MeV}$.  The ratio of momentum transfer and Fermi
  momentum is kept fixed at $q/k_F=0.01$ and the Fermi momentum is set
  to $k_F=1.0\,\textrm{fm}^{-1}$, which translates to the density
  $n=0.221n_0$. }
\label{fig:reimpi}
\end{center}
\end{figure}

\begin{figure}[tbh]
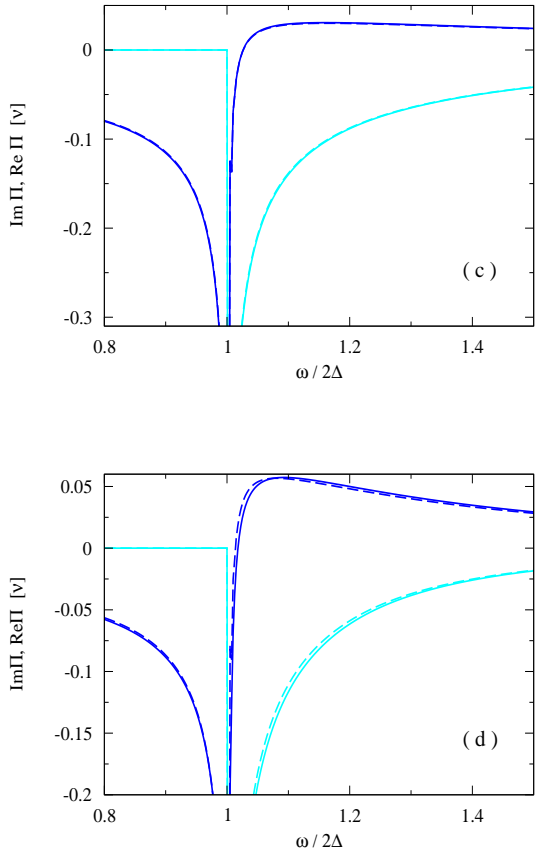

\begin{center}
\includegraphics[width=7.0cm,height=5.0cm,angle=0]{spincurrent_kf10_05tc.eps}
\vskip 1.2cm
\includegraphics[width=7.0cm,height=5.0cm,angle=0]{spindensity_kf10_05tc.eps}
\end{center}
\caption[] {(Color online) The same as in Fig.~\ref{fig:reimpi}, 
but for (c) the spin-current response function 
$\Pi^S_{00}$  and (d) for the spin-density response function
$\Pi^S_{jj}$ (d). }
\label{spin_v1}
\end{figure}

Figures~\ref{fig:reimpi} and \ref{spin_v1}
show the dependence of the real and imaginary
parts of the density and spin response functions, respectively, 
 on the transferred energy for
fixed three-momentum transfer.  The zero temperature gap is fixed at
$\Delta (0) = 1$ MeV and $T_c =\Delta(0)/1.76$.  The lowest order
Landau parameter is set $v_{\ph}^D = -0.5$ for density perturbations
and $v_{\ph}^S = 0.5$ for spin perturbations (these correspond to the
values computed in Ref.~\cite{Sedrakian:2006ys}).  The frequency and
momentum transfer are normalized to the threshold frequency
$2\Delta(T)$.  The response function in the negative energy range can
be obtained from the relations ${\rm Re}\Pi^{D/S}(-\omega) = {\rm
  Re}\Pi^{D/S}(\omega)$ and ${\rm Im}\Pi^{D/S}(-\omega) = -{\rm
  Im}\Pi^{D/S}(\omega)$.  The numerical method of computing the
response functions exactly is described in Appendix~\ref{AppendixB}.
\begin{figure}[!]
\begin{center}
\includegraphics[width=6.5cm,angle=-90]{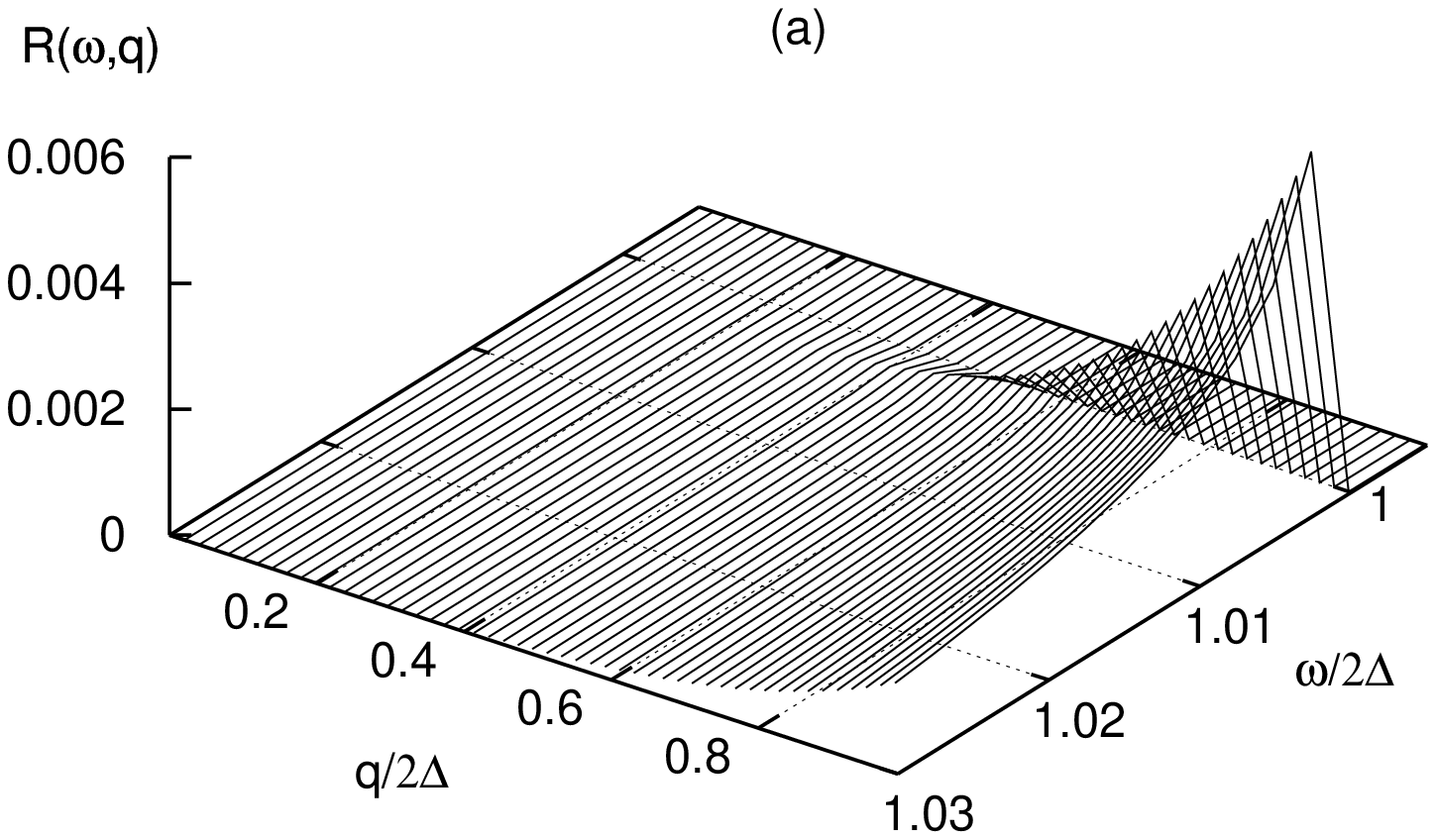}
\includegraphics[width=6.5cm,angle=-90]{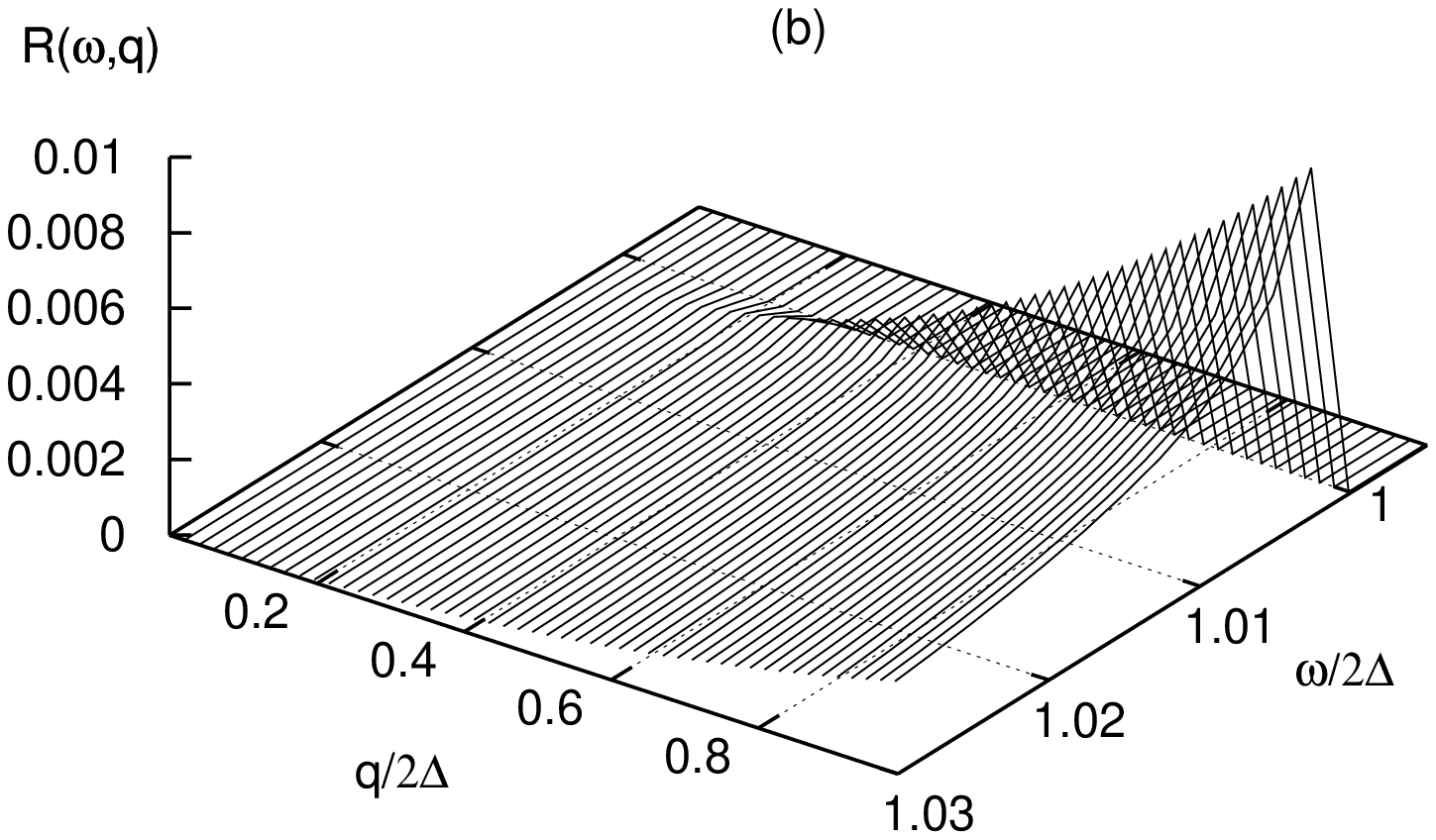}
\includegraphics[width=6.5cm,angle=-90]{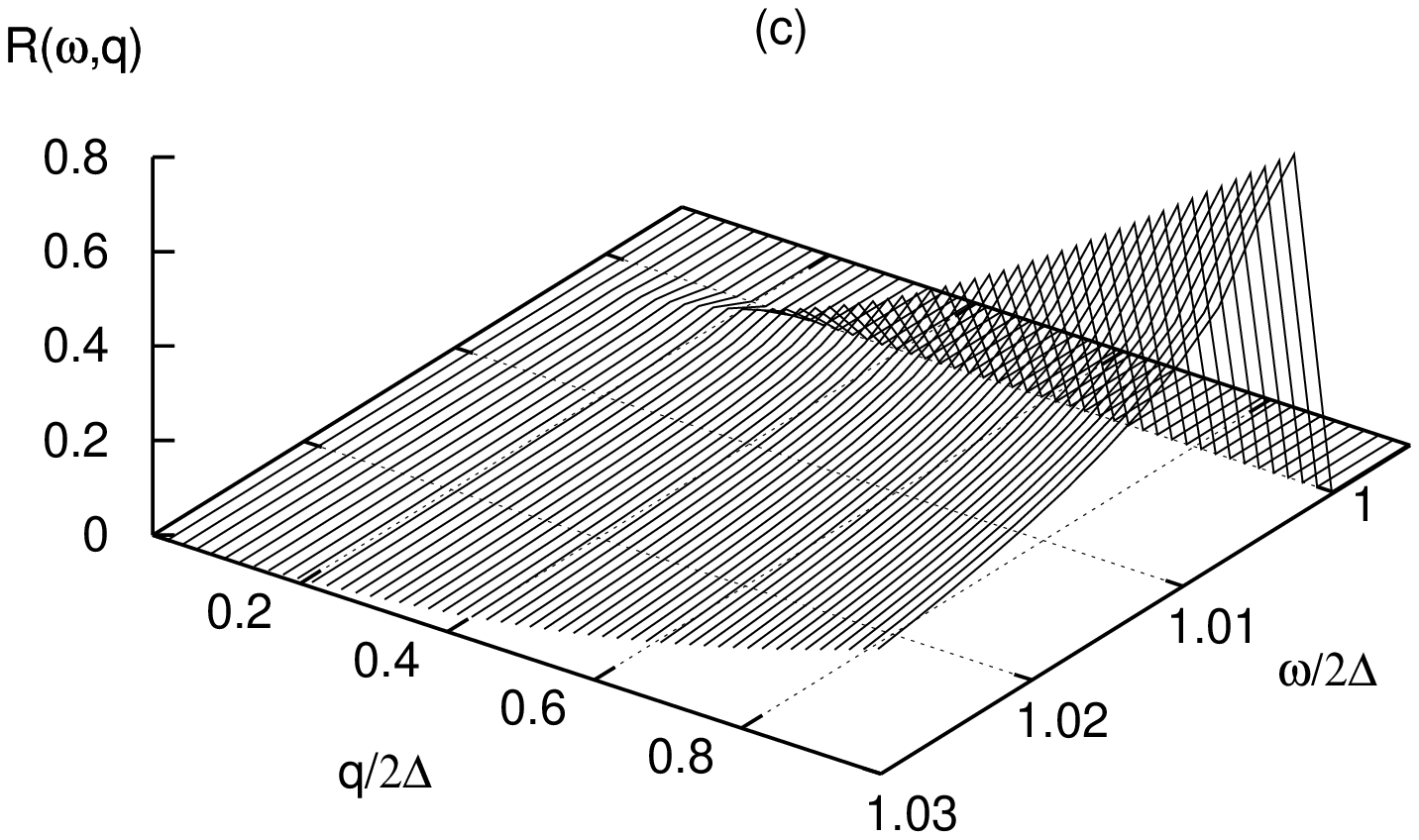}
\end{center}
\caption[] {(Color online)
The spectral function of (a) the density fluctuation, (b) current
fluctuations and (c) spin-density fluctuations as a
 function of the energy transfer $\omega$ and of momentum transfer in
 units of $2\Delta(T)$ at  $T = 0.5 T_c$ and $k_F = 1$ fm$^{-1}$,
 which corresponds to density $0.211\,n_0$, where 
 $n_0=0.16~$fm$^{-3}$ is the nuclear saturation density. 
}
\label{spfunc_den}
\end{figure}

Our comparison of the perturbative analytical results with the exact
numerical ones shows that (i)~for the density response the 
higher-order corrections shift the imaginary part to higher frequencies, \ie,
for a fixed frequency the imaginary part is larger; the real parts are
correspondingly larger as well. (ii)~For the density current response
the perturbative and exact results match to a high accuracy; (iii)~for
the spin-current response both results match again to a high accuracy;
(iv) for the spin-density response small deviations are observed close
to the threshold; the imaginary part is again shifted to higher
frequencies.  Note that in each case the imaginary parts are
identically zero below the threshold for pair breaking process
$2\Delta(T)$.

The density response function can be compared to the one derived in
a previous paper~\cite{Sedrakian:2010xe}.  As shown above, the first
non-vanishing contribution arises from the term $\Pi_{D4}^{00}$ and
not from $\Pi_{D2}^{00}$ as in Ref.~\cite{Sedrakian:2010xe}, where
$\Pi_{D2}^{00}\neq 0$.  Consequently, the numerical values of the real
and imaginary parts are roughly by an order of magnitude
smaller. Nevertheless, the dependence of the real and imaginary parts
of the polarization tensor on the frequency shows essentially the same
behavior.  The difference between the present results and that of 
Ref.~\cite{Sedrakian:2010xe} can be understood as follows. We note that the
general form of the density response function in
\cite{Sedrakian:2010xe}, Eq. (18) and the definitions of the
elementary loops, Eqs. (19) to (22), are the same. The difference arises
at the level of the loops ${\cal A}, {\cal B},$ and ${\cal C}$ given
by Eq. (25) to (27) of \cite{Sedrakian:2010xe}. 
The most general form of the first loop, upon substitution of
Bogolyubov amplitudes in Eqs. (19) and (21) of Ref.~\cite{Sedrakian:2010xe} is given by 
 \be\label{B1} {\cal A}(q) = \int\!\!\!
\frac{d^3p}{(2\pi)^3} \left[(\ep+\ep') (\ep\ep'-\xi\xi'+\Delta^2)
  +\omega (\xi'\ep-\xi\ep') \right]\mathscr{G} 
\ee
 with the short-hand
notations $\xi = \xi_{\vecp-\vecq/2}$, $\xi' = \xi_{\vecp+\vecq/2}$
and 
$\ep = \sqrt{\xi^2+\Delta^2}$, $\ep' = \sqrt{(\xi')^2+\Delta^2}$ and 
\bea \label{B2} \mathscr{G} &=& \frac{1}{2\ep\ep'}
\left[\frac{1-f(\ep)-f(\ep')}{\omega^2-(\ep+\ep')^2}\right].
\eea
We see that  Eq. (25) of Ref.~\cite{Sedrakian:2010xe} does not contain
the $\omega (\xi'\ep-\xi\ep')$ which vanishes manifestly  in the limit
$\vecq \to 0$, but is finite if $\vecq \neq 0$.
For the remaining ${\cal B}$ and ${\cal C}$ loops we obtain 
\bea
\label{B3} {\cal B}(q) &=& 2\Delta \int\!\!\!\frac{d^3p}{(2\pi)^3}
\left[\omega \ep' + (\ep'+\ep)\xi' \right]\mathscr{G}\\
 {\cal C}(q) &=& \int\!\!\!\frac{d^3p}{(2\pi)^3}
\Biggl\{ \frac{1-2f(\ep)}{2\ep}
- \Bigl[(\ep+\ep')(\ep\ep'+\xi\xi'+\Delta^2)\nonumber\\
& -& \omega (\xi\ep'+\xi'\ep)
\Bigr]\mathscr{G}\Biggr\}
\eea
and we see that the terms 
$(\ep'+\ep)\xi' $ and $-\omega (\xi\ep'+\xi'\ep)$ are missing in
Eqs.~(26) and (27) of Ref.~\cite{Sedrakian:2010xe}.
 Both terms that were dropped vanish in the limit $\vecq \to 0$,
because they are odd in $\xi$, while after changing the integration
measure according to Eq. (\ref{intmeasure}), we obtain integrals 
over symmetrical in $\xi$ limits. Thus, we conclude that, the discrepancy
between the present treatment and that of Ref.~\cite{Sedrakian:2010xe}
originates from incomplete expressions in Eqs. (25)-(27) of the latter work.

\section{Spectral functions and collective modes}
\label{sec:5}
The knowledge of the response functions allows us to construct an
effective theory of excitations in the nuclear medium. Their full
(interacting) propagator is completely determined by their spectral
function, which in each channel is defined via  the imaginary part of
the polarization as
$R(\omega,\vecq) =-2{\rm Im}\Pi(\omega,\vecq)$. For
example, in the density channel, using Eq.~(\ref{eq:res_D00}),  one finds 
\bea 
R(\omega,\vecq)= -\frac{2 (v_{\ph}^D)^{-2}{\rm Im} {P}(\omega,\vecq)}
    {\left[(v_{\ph}^D)^{-1} - {\rm Re} {P}(\omega,\vecq) \right]^2 
      + {\rm Im}{P}(\omega,\vecq)^2}, \nonumber\\
\eea
where ${P} (\omega,\vecq)=
\mathcal{Q}^+(\omega,\vecq)/\mathcal{C}(\omega,\vecq).$ Similar
relations hold for other  excitation channels (\ie, 
current-density, spin-current and spin-density). Above the threshold
$\omega > 2\Delta(T)$ non-zero imaginary part implies that the
collective excitations have finite life-time, \ie, are not perfect
quasiparticles. Nevertheless, in the limit where the imaginary part is
small one can approximate the spectral function as
\be
\label{S_SMALL} 
R(\omega,\vecq) = 
2\pi Z(\vecq) \delta(
(v_{\ph}^D)^{-1} - {\rm Re} {P}(\omega,\vecq)
) + R_{\rm reg}(\omega,\vecq), \nonumber\\
\ee 
where $R_{\rm reg}(\omega,\vecq)$ is the regular (\ie ~smooth) part 
of the spectral functions and $Z(\vecq)$ is the wave-function renormalization.
The dispersion relation of
the excitations is given by the solution $\omega(\vecq)$ of the equation
\be \label{eq:QP_spectrum} 
1-v_{\ph}^D{\rm Re}{P}(\omega,\vecq) = 0.
\ee 
Figure~\ref{spfunc_den} shows the dependence of the spectral functions
for density [Fig.~\ref{spfunc_den}(a)], current
[Fig.~\ref{spfunc_den}(b)],
 and spin-density  [Fig.~\ref{spfunc_den}(c)] fluctuations on the
energy and momentum transfer.  The spectral functions have a
Breit-Wigner form, therefore the location of their maxima is
controlled by the real parts of the response functions, whereas their
widths by the imaginary parts.  In the case of density fluctuations
the imaginary component of the polarization tensor has a power-law
($\propto q^4$) behavior for fixed energy transfer, as is explicit
from the analytical form (\ref{eq:pi_density}).  At fixed momentum
transfer the spectral function has a threshold due to the
proportionality ${\cal G}_0\propto \theta (\omega -2\Delta)$. At
low-momentum transfers the main contribution to the spectral function
comes from the vicinity of the pair breaking threshold ($\omega \sim
2\Delta)$; for large momentum transfers, modes away from the energy
threshold become important.  The qualitative features seen in the
spectral function of the density response are seen also for the
current response; some quantitive differences arise because now, for
fixed energy transfer, the imaginary part scales as $\propto q^2$,
c.f. Eq.~(\ref{eq:pi_current}).  Consequently, the low-momentum
contributions are only weakly suppressed and the maximum of the
spectral function is numerically larger.  The response functions
associated with spin perturbations appear at order $v_F^2$, therefore
their absolute scale is larger than that for the density and
current-density responses, which scale as $v_F^4$.  It has the same
functional dependence on the momentum and energy transfer as the
density-current response [see Eq.~(\ref{eq:pi_spin})] and differs only
by the numerical pre-factor and the $v_F^2$ dependence.  For small
momentum transfers the main contribution to the spectral function
comes from the region near the threshold.  Note that the spin-current
response, to leading order, is independent of the momentum transfer,
therefore the two-dimensional form given in Fig.~\ref{spin_v1} (c)
is sufficient.  Its dependence on the frequency reflects the
dependence of the function ${\cal G}_0$,
c.f. Eq.~(\ref{eq:pi_spincurrent}).
\begin{figure*}[hbt]
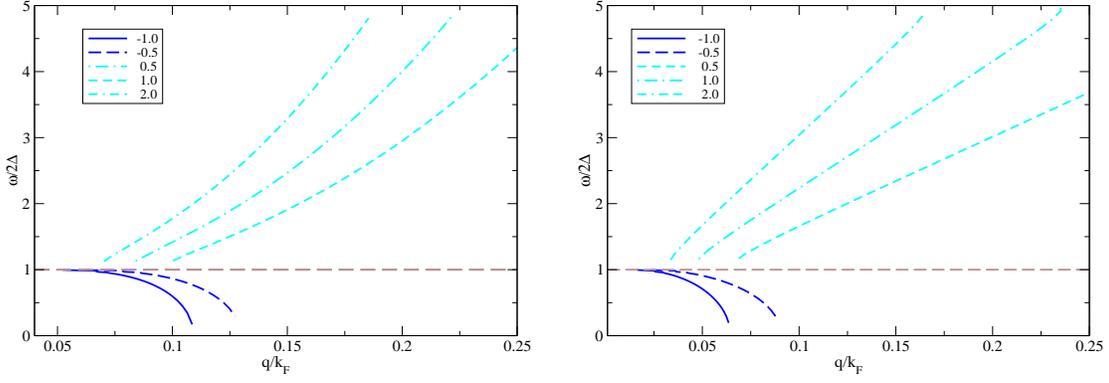

\hskip 1cm
\begin{center}
\includegraphics[width=7.0cm,height=5.0cm,angle=0]{densexitons.eps}
\hskip  0.5cm
\includegraphics[width=7.0cm,height=5.0cm,angle=0]{spinexcitions.eps}
\end{center}
\caption[] {(Color online)
  Dispersion relations of collective excitations for density
   (left panel) and spin density (right panel) perturbations for
   the values of the particle-hole interaction $v_{\ph}$ shown in 
   each panel for $k_F = 1$ fm$^{-1}$ and $T/T_c =0.5 $. 
   The heavy lines (blue online) correspond to
   undamped exitonic modes, the light  lines (cyan online)
   correspond to diffusive damped pair-breaking modes. 
}\label{fig_dispersion}
\end{figure*}

From the spectral functions we can extract the
quasiparticle spectra of the collective excitations. These can be
defined by the poles of the spectral function when
${\rm Im}{P}(\omega,\vecq) = 0$, \ie,  by the condition
(\ref{eq:QP_spectrum}).
We start by setting the parameters characterizing the superfluid state
to their relevant scales and by studying the nature of the modes as a
function of the particle-hole interaction $v_{\ph}$. The stability of
the normal Fermi-liquid state constrains $v_{\ph} > -1$ (note that we 
work at leading order in the expansion of Landau parameters in 
spherical harmonics).  The numerical
solutions of Eq. (\ref{eq:QP_spectrum}) for the density and spin
excitations are shown in Fig.~\ref{fig_dispersion} for
$-1 \le v_{\ph}\le 2$ and fixed $k_F = 1$ fm$^{-1}$, $T/T_c =
0.5$ with $\Delta = 1$ MeV.  For positive values of the particle-hole
interaction the modes appear in the domain $\omega/2\Delta > 1$, where
${\rm Im}{P}(\omega,\vecq) \neq 0$, \ie, they represent damped
(diffusive) modes of oscillations of density and spin-density,
respectively, associated with the pair-breaking processes. For
negative values of the particle-hole interaction, the modes exist in
the domain $\omega/2\Delta \le 1$, where the pair-breaking part of the
${\rm Im}{P}(\omega,\vecq)$ vanishes; therefore the modes
represent undamped oscillations of density and spin-density around
their average values. These modes are ``exitonic" as they correspond
to bound pairs of particles and holes.
\begin{table}[tbh]\caption{The values of the coefficients in the fit
    formula (\ref{eq:fit}) for several values of the particle-hole
    interaction in the density (upper part) and spin (lower part)
    channels. The remaining parameters are fixed to their
    characteristic values $k_F = 1$ fm$^{-1}$, $\Delta = 1$ MeV, and $m^*/m
    =1$. The interactions
are given in units of the density of states $\nu(k_F)$.}
\begin{tabular}{rrrrr}
\hline
$v_{\ph}$ & $a$ &  $b$ &  $c$ & $d$ \\
\hline
$-1~$    & 0.817879 & 88.4029 & $-$ 11411.2 & \\
$-0.5~$  & 0.849483 & 53.7703 & $-$ 5210.88 &\\
0.5~   & 0.600277 & 80.4847 &  115.199 &\\
1~    & 0.586293 & 57.3376 & 49.0417 &\\
2~     & 0.584029 & 116.109 &  193.221 &\\
\hline
$-1~$   & 1.02616  & $- 84.0295$  &  2216.43  & $-7449470.$ \\
$-0.5~$ & 1.02487  & $- 39.6023$  & $- 568.605$  & $- 844334$. \\
0.5~  & 0.885174  & 83.0049  & $-$ 922.6   & 4886.17 \\
1~    & 0.928416 &  154.6  & $- 2912.19$   & 26000.1 \\
2~    &  0.924387 &  313.859  & $- 12031.8$   &  218234. \\
\hline
\end{tabular}
\label{tab:1}
\end{table}
\begin{table}
  \caption{The same as in Table~\ref{tab:1}, but for the
    density-dependent $v_{\ph}$ interactions 
    taken from Ref.~\cite{Sedrakian:2006ys}.
    The entries are the Fermi wave vector, 
    the  values of the coefficients
    in the fit formula (\ref{eq:fit}), and the range of momentum transfers
    $\Delta q$ in units of $k_F$.
  }
\begin{tabular}{llllll}
\hline
$k_F$&$v_{\ph}$&$a$&$b$&$c$&$k_F^{-1}\Delta q $\\
 &                &       &         &
&\\
\hline
\hline
 1.0   &   $
\begin{array}{c} -0.45      \\ 0.41 \end{array} 
$  & $
\begin{array}{c}  0.941366    \\  0.518545 \end{array}
$& $
\begin{array}{c}  2.69439  \\  12.4875  \end{array}
$ &$
\begin{array}{c} - 34.0444 \\  8.8972 \end{array} 
$&$
\begin{array}{c}   ( 0.197;0.29)               \\    (0.223 ;0.3)          \end{array}    $
\\
\hline
1.2   &$
\begin{array}{c} -0.43     \\  0.40 \end{array} 
$   &  $
\begin{array}{c} 0.896992     \\0.6465 \end{array}   
$ & $ 
\begin{array}{c} 12.9105 \\ 37.092 \end{array}  
$ & $
\begin{array}{c} - 447.747\\ - 197.621\end{array}
$ & $ 
\begin{array}{c}   (0.111; 0.22)\\   (0.127;0.3)   \end{array}    
$
\\
\hline
1.4   & $ 
\begin{array}{c} -0.41    \\ 0.40 \end{array} $   & $
\begin{array}{c} 0.901629 \\  1.17054 \end{array} $& $
\begin{array}{c} 67.8547\\  84.3504\end{array} $& $
\begin{array}{c} - 12952.2\\  - 488.556\end{array}  $&$
\begin{array}{c}   (0.047; 0.098)  \\     (0.054;0.3)          \end{array}    $\\
\hline
 1.6    &  $ 
\begin{array}{c}  -0.36      \\ 0.39 \end{array}  $&$
\begin{array}{c} 0.894476 \\  1.15211  \end{array}$ &$
\begin{array}{c} 702.973\\  859.001\end{array} $ & $
\begin{array}{c} - 1283650\\  - 49727.6\end{array} $&$
\begin{array}{c}     (0.015; 0.031)    \\     (0.018;0.08)        \end{array}    $\\
\hline
\hline
\end{tabular}
\label{tab:2}
\end{table}
The spectra in each case can be accurately fitted by the polynomial of
the form
\be \label{eq:fit}
\tilde \omega (q) = a + b \tilde q^2 + c\tilde q^4 + d \tilde q^6, 
\ee
where in the case of density perturbations accurate results are
obtained with only three parameters  ($d=0$). Here we defined
dimensionless quantities $\tilde \omega = \omega/2\Delta$ and $\tilde
q = q/k_F$.
The fitted values of the parameters for the results shown in
Fig.~\ref{fig_dispersion}  and are given in Table~\ref{tab:1}. 

Next we consider a specific microscopic
calculation~\cite{Sedrakian:2006ys}, which provides us with the
density dependence of the parameters of the neutron superfluid {\it and }
the associated values of the leading-order Landau parameters in the
particle-hole channel. We  solved Eq. (\ref{eq:QP_spectrum}) in
the density and spin channels for each density and subsequently fitted
the spectra with the formula (\ref{eq:fit}). The results are displayed
in Table~\ref{tab:2}; some of the characteristics of the superfluid
are shown in Table~\ref{tab:3}. The density excitations exist below
the pair-breaking threshold, \ie, represent undamped exitonic
modes. Conversely, because $v_{\ph}$ changes the sign in the spin
channel, the spin excitations represent diffusive modes with finite
damping. Note that each of these modes exist within some finite
interval $\Delta q$ of momentum transfers.  The lower bound arises
because perturbations that are sufficiently large to excite a mode
arise at some {\it finite value} of $q$. The upper bound in most cases
is the consequence of the use of perturbative response functions,
whose validity breaks down for large momentum transfers $q/k_F \sim
0.3$; in some cases the upper bounds are associated with the
disappearance of the solutions from the search domain.

\section{Specific heat}
\label{sec:6}

The specific heat contribution arising from the collective modes in the
neutron star crust has recently attracted recently attention in the context of
non-spherical phases~\cite{DiGallo:2011cr}. These modes at not too low
temperatures can dominate the specific heat provided by the
degenerate, ultra-relativistic electron gas. Below we shall examine
the contribution of the collective modes discussed in the previous
section to the specific heat of a superfluid neutron star crust.

The entropy of a collective bosonic mode is given by
\bea\label{eq:entropy} S = 2k_B \sum_\vecq \left[ (1+g_\vecq)\ln
  (1+g_\vecq) + g_\vecq \ln g_\vecq\right], \eea where $g_\vecq =
[\exp(\omega_\vecq/T)-1]^{-1}$ is the Bose distribution function of
collective excitations with the spectrum $\omega_\vecq$. The specific
heat is then given by \bea \label{eq:specheat1} c_V = k_BT
\frac{dS}{dT} = k_B \sum_\vecq\frac{\omega_\vecq}{T}\,\frac{\partial
  g_\vecq}{\partial T}.  \eea For a collective (acoustic) mode with
linear spectrum $\omega = u \vert \vecq \vert$, where $u$ is the sound
velocity, Eq.~(\ref{eq:specheat1}) can be integrated~\cite{Ashcroft}
\be \label{eq:cv_a} c_V^{(a)} = \frac{2\pi^2 k_B^4 T^3}{15 (\hbar
  u)^3}.  \ee An acoustic mode, in a compact star setting, is associated
with the nuclear lattice in the crust, where phonons contribute to the
specific heat below the melting temperature of the crust $\sim
10^9$~K.  At low temperatures the superfluid supports the
Bogolyubov-Anderson (BA) mode with the velocity 
\be \label{eq:AB}
u^{\rm BA} = \frac{v_F}{\sqrt{3}}(1+v^D_{\ph})^{1/2}.  
\ee 
where $v_F =
\hbar k_F /m^*$ is the (effective) Fermi velocity.  The dispersion
relation (\ref{eq:AB}) does not contain temperature corrections. In
the following we will ignore the damping of the BA mode and
extrapolate the result (\ref{eq:AB}) to higher temperatures.  Apart
from these two collective modes, the main contribution to the specific
heat of matter is due to the electrons which, in a first
approximation, can be treated as a uniform ultra-relativistic ideal
Fermi gas. At low temperatures their specific heat is then given by
\be\label{eq:cv_e} c_V^{(e)} = \frac{k_B^2 \mu_e^2 T}{3(\hbar c)^3},
\ee where $\mu_e$ is the electron chemical potential.

Table \ref{tab:3} compares the various contributions to the specific
heat of matter at subnuclear densities.  The temperature at each
density corresponds to $T = 0.5 T_c$, with $T_c = \Delta/1.76$.  The
contribution of the BA mode, $c_V^{\rm BA}$ is computed
from Eqs. (\ref{eq:cv_a}) and (\ref{eq:AB}), the contribution of
electrons from Eq. (\ref{eq:cv_e}) assuming $n_e = 0.023 n_n$, where
$n_e$ and $n_n$ are the electron and neutron number densities.  The
contributions from density and spin pair-breaking contributions,
$c_V^{(\rho)}$ and $c_V^{(\sigma)}$ are computed through the numerical
integration of Eq.~(\ref{eq:specheat1}) with the collective mode
spectrum given by Eq.~(\ref{eq:fit}). The coefficients $a$, $b$ and
$c$ in Eq.~(\ref{eq:fit}) for the density fluctuations and the spin
fluctuations, as well as the integration limits in
Eq.~(\ref{eq:specheat1}) are tabulated in Table \ref{tab:2} (the
coefficient $d = 0$ in all cases). It is seen that the density
fluctuations considerably contribute to the net specific heat of
matter for lower densities (wave-vectors), while the spin-fluctuations
are negligible at $T/T_c = 0.5$. The result of for the BA mode should
be taken as suggestive, because we neglected the temperature
correction to the dispersion relation and the possible damping of this mode.

\begin{figure}[tbh]
\hskip 1cm
\begin{center}
\includegraphics[width=7.3cm,height=5.0cm,angle=0]{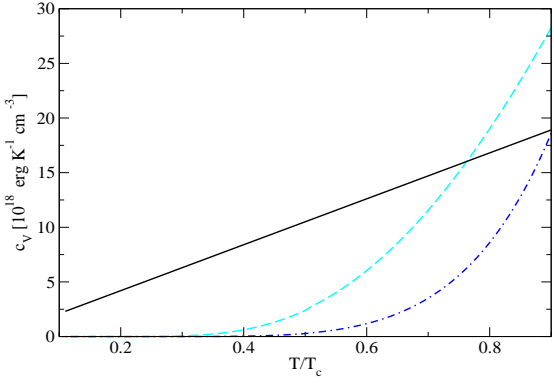}
\end{center}
\caption[] {(Color online)
Dependence of the specific heats due to electrons (solid line) 
 and pair-breaking
density (dashed line) and spin fluctuations  (dash-dotted line) 
on the reduced temperature $T/T_c$ for $k_F = 1$ fm$^{-1}$.
}\label{fig:cv}
\end{figure}

The temperature dependence of the specific heat due to the
pair-breaking modes and the specific heat of electron gas is shown in
Fig.~\ref{fig:cv} for $k_F = 1 \,$fm$^{-1}$.  The electron specific
heat is linear in temperature, whereas the specific heat of the
pair-breaking fluctuations has a power law behavior, which is close to the
$T^3$ law characteristic for linear in $q$ spectra. The difference
reflects the non-linearity of the spectrum (\ref{eq:fit}). We have
assumed that the temperature dependence of the coefficients $a$, $b$,
and $c$ can be neglected in a first approximation, \ie, the spectrum
of collective excitations is assumed to be independent of temperature.
This assumption is validated by the insensitivity of the maxima of the
spectral functions to the temperature variations (see
Fig.~\ref{spfunc_den}) which were compared at $T/T_c = 0.5$ and 0.9.

\begin{table}
\caption{ 
  Density and wave-vector dependence of the specific heat of
  various components in neutron matter and the crust of a neutron
  star at $T = 0.5 T_c$. Tabulated are the net density of matter
  $n_n[\, $fm$^{-3}]$, the neutron wave-vector $k_F[$fm$^{-1}]$, 
  the neutron effective mass in units of bare mass, the pairing gap 
  $\Delta [$MeV$]$, the electron wave-vector $k_{Fe}[$fm$^{-1}]$,
  assuming that the electron number density $n_e = 0.023n_n$,  the
  specific heats of electron gas $c_V^{(e)}$,  Bogolyubov-Anderson
  phonons $c_V^{(\rm BA)}$, pair-breaking density fluctuations
  $c_V^{(\rho)}$, pair-breaking spin fluctuations $c_V^{(\sigma)}$ in 
   units of $10^{18}$ erg cm$^{-3}$ K$^{-1}$.
}
\begin{tabular}{ccccccccc}
\hline 
$n_n$ & $k_{F}$  & $m^*$ & $\Delta$ & $k_{Fe}$   & $c_V^{(e)}$  &
$c_V^{(\rm BA)}$ & $c_V^{(\rho)}$& $c_V^{(\sigma)}$ \\
\hline
0.034 & 1.00 & 0.94 & 3.09 &  0.29 &   16.6  & 26.419 & 2.479 & 0.291  \\
0.058 & 1.20 & 0.92& 2.44 &   3.42&   18.9  & 7.611 & 3.508 & 0.079  \\
0.093 & 1.40 & 0.88 & 1.41 &  3.99 &   14.8  & 1.003 & 0.008 & 0.004 \\
0.138 & 1.60 & 0.84& 0.57 &   0.45  &    7.8  & 0.045 & 0.012 & 0.000 \\
\hline
\end{tabular}
\label{tab:3}
\end{table}

\section{Conclusions}
\label{sec:7}

In this work we studied the response functions of a single component
pair-correlated baryonic matter to density, spin and their current
perturbations in the low-temperature regime. These results should be
relevant for the description of both the dynamical and thermodynamical
properties of baryonic matter at low densities, \ie, the densities
where the baryons form an $S$-wave superfluid. It was observed that
the expansions in the parameters $q/k_F$ and $v_Fq/\omega$ lead
essentially to the same perturbative results, which in turn can be
interpreted as an expansion in the parameter $v_F\ll 1$. We have
applied an exact numerical method to evaluate the response functions
and to validate the perturbative approximation in the domain of its
convergence.  We further derived the dispersion relations of the
collective excitations of density and spin-density perturbations. For
positive values of the particle-hole interactions these correspond to
weakly damped diffusive excitations, whereas for negative values - to
undamped excitonic modes.

The spectral functions presented above can be modified in a number
of ways. As noted in the Introduction the multi-loop processes were
found to be important for the neutrino emission and they could
additionally contribute to the spectral functions in the kinematical
domain where two-particle-two-hole excitations are
important. Furthermore, higher order Landau parameters, if included
into driving interactions in the particle-particle and particle-hole
channels may require some renormalization of the spectra,
see~Refs.~\cite{Vaks,Kolomeitsev:2011wz,Keller_Thesis,Voskresensky:1987hm}.

The application of the formalism to compute the specific heat of the
matter expected in neutron star crusts shows that the contribution of
the collective pair-breaking excitations can be a significant part of
the net specific heat of matter. For some density parameters and not
too low temperatures the combined contribution from superfluid modes
of neutron fluid can be larger than the specific heat stored in the
degenerate electron gas.

\section*{Acknowledgment}

This work was supported by the Deutsche Forschungsgemeinschaft 
Grant No. SE 1836/1-2 (JK), the HGS-HIRe graduate program (JK), and 
by GSI (AS).

\appendix

\section{Solving the equations for the vertices}
\label{AppendixA}
The bare vertices given by Eq.~(\ref{eq:vertices2})
are diagonal in spin space. Likewise, the particle and hole vertices
are diagonal in spin space, \ie, $\hat{\Gamma}_1=\Gamma_1\hat{1}_2$
and $\hat{\Gamma}_4=\Gamma_4\hat{1}_2$.  The anomalous vertices are
proportional to the second Pauli matrix, $\hat{\Gamma}_2=\Gamma_2
i\sigma_2$ and $\hat{\Gamma}_3=\Gamma_3 i\sigma_2$.  The equation for
the hole vertex $\Gamma_4$ can be obtained from the equation for the
particle vertex $\Gamma_1$ by interchanging particle and hole
lines. To account for this property one can introduce,
following Ref.~\cite{Migdal:1967}, an operator $\hat{\mathcal{P}}$ to
revert the direction of ingoing and outgoing momenta and to exchange the
spin indices simultaneously, when acting on a vertex function. The
explicit action of this operator is \be
\label{eq:rel1}
{\cal P}\Gamma_{4,\alpha\beta} 
=\sigma_y\Gamma^*_{4,\alpha\beta}\sigma_y ={\cal
  T}\Gamma_{1,\alpha\beta}, \ee whereby ${\cal T}$ is the
time-reversal operator, \ie, it is equal to $+1$ for vertices which
are even under time reversal operation and $-1$ for vertices which are
odd under this transformation.  By considering the action of the
operator $\hat{\mathcal{P}}$ on the bare vertices one finds that
scalar vertices, \eg,  $\Gamma_0^D = 1$ and 
$\Gamma_0^S = {\bm{\sigma}}\vecv$
do not change their sign, while vector vertices like
${\bm{\Gamma}}_0^D = \vecv$ or ${\bm{\Gamma}}_0^S = {\bm{\sigma}}$
gain an additional minus sign.  Furthermore, we note that the
equations for $\Gamma_2$ and $\Gamma_3$ are adjoint to each
other. Formally, one can cast this property into the equation \be
\label{eq:rel2}
\hat{\tilde{\Gamma}} = -{\hat \Gamma}_2 = {\hat \Gamma_3}^+.
\ee
Note that the full current vertices can depend on any external momentum
involved in the problem, therefore they need to be decomposed in
components along the vectors $\vecv$ and $\vecq$.  Thus, the most
general Ansatz for the density current vertices is 
\bea\label{eq:decomp1}
\bm{\Gamma}^{D} &=& \Gamma^{D}_{v}\vecn_{v}+\Gamma^{D}_{q}\vecn_{q},\\
\label{eq:decomp2}
\tilde{\bm{\Gamma}}^{D} &=& \tilde{\Gamma}^{D}_q\vecn_{q}, \eea where
the subscripts on the unit vector $\vecn$ refer to the vector
defining its direction. The coefficients $\Gamma^{D}_{v,q}$ are
normalized such that $\Gamma^{D}\vecn_v = \vecv_F$, \ie, the
coefficient $\Gamma^{D}$ is simply the modulus of the Fermi velocity.
Similar to Eqs.~(\ref{eq:decomp1}) and (\ref{eq:decomp2})
decompositions holds for spin-current vertices.

The solution of the system (\ref{G1}) to (\ref{G4}) is simplified if one
takes into account the  identities~\cite{Migdal:1967,Leggett:1966zz} 
\bea
GG^-(\vecv,\omega,\vecq) &=& G^-G(\vecv,\omega,\vecq),\\
G^-F(\vecv,\omega,\vecq) &=& -FG^-(\vecv,\omega,\vecq),\\
GF(\vecv,\omega,\vecq) &=& -FG(\vecv,\omega,\vecq),\\
G^-G^-(\vecv,\omega,\vecq) &=& GG(\vecv,-\omega,\vecq),
\eea
where the products of the Green's functions refer to their convolutions
defined as z
\bea        
X_+ X'_-&=&\nu T\sum\limits_{n=-\infty}^{\infty}\int\limits_{-\infty}^{\infty}\!d\xi_p\,
X\left(ip_n+i\omega_m,\vecp+\frac{\vecq}{2}\right)\nonumber\\
&\times&X'\left(ip_n,\vecp-\frac{\vecq}{2}\right),
\eea
where $X_{\pm} \in \{G_\pm, G^-_\pm, F_\pm, F^+_\pm\}$. 
The solution contains the following 
linear combinations of the 
convolutions~\cite{Abrikosov:1962,Migdal:1967,Leggett:1966zz}:
\bea
\label{eq:loopA}
A(\hat{\mathcal{P}})&=&G_+G_- -F_+F_-\hat{\mathcal{P}},\\
\label{eq:loopB}
B&=&G_+F_- -F_+G_-,\\
\label{eq:loopC}
C&=&G_+G_-^- +F_+F_- -v_{\pp}^{-1},\\
\label{eq:loopD}
D(\hat{\mathcal{P}})&=&-G_+F_- -F_+G_-^-\hat{\mathcal{P}}.
\eea
We take the matrix structure of vertices and propagators into account and use the 
relations (\ref{eq:rel1}) and (\ref{eq:rel2}) to cast the set of the 
four coupled integral equations into the following two equations for 
the new vertex functions
\bea
\label{eq:Gamma1}
\Gamma^{D/S}(\vecv,\omega,\vecq)
&=& \Gamma_0^{D/S}(\vecv,\omega,\vecq)
   +\int\!\frac{d\Omega'}{4\pi}\,V_{\ph}^{D/S}(\vecv,\vecv')\nonumber\\
 &&\hspace{-2cm}\times \left[A(\hat{\mathcal{P}})\Gamma^{D/S}(\vecv',\omega,\vecq)
   +B\tilde{\Gamma}^{D/S}(\vecv',\omega,\vecq)\right],\\
\label{eq:Gamma2}
\tilde{\Gamma}^{D/S}(\vecv',\omega,\vecq)
&=& \int\!\frac{d\Omega'}{4\pi}\,V_{\pp}^{D/S}(\vecv,\vecv')\nonumber\\
 &&\hspace{-3cm}\times \left[(C+v_{\pp}^{-1})\tilde{\Gamma}^{D/S}(\vecv',\omega,\vecq)
   +D(\hat{\mathcal{P}})\Gamma^{D/S}(\vecv',\omega,\vecq)\right].\nonumber\\
\eea
To write down the solutions of the integral equation we need the
following angle averages of the loop functions
\bea
\mathcal{A}(\hat{\mathcal{P}})
&=&\int\frac{d\Omega}{4\pi} ~A(\hat{\cal P}),\\
\mathcal{B}
&=&\int\frac{d\Omega}{4\pi} ~B,\\
\mathcal{C}
&=&\int\frac{d\Omega}{4\pi} ~C,\\
\mathcal{D}(\hat{\mathcal{P}})
&=&\int\frac{d\Omega}{4\pi} ~D(\hat{\cal P}),
\eea
where $d{\Omega} = \sin\theta\,d\theta\,d\phi$. Furthermore, we 
need the angle averages of first moments of the loop functions
with respect  to the cosine of the
angle enclosed by $\vecn_v$ and $\vecn_q$, \ie, 
$x \equiv \vecn_{\vecq}\cdot \vecn_{\vecv}$, which we write as
\be
\tilde{\mathcal{Y}}=\int\frac{d\Omega}{4\pi}\,\tilde{Y},
\ee
where 
\bea
\tilde{Y} &=& x\,\nu T\sum\limits_{n=-\infty}^{\infty}\int\limits_{-\infty}^{\infty}\!d\xi_p\,
         X\left(ip_n+i\omega_m,\vecp+\frac{\vecq}{2}\right)\nonumber\\
&\times &X'\left(ip_n,\vecp-\frac{\vecq}{2}\right).
\eea
We also define the auxiliary combination of the loops:
\bea 
\mathcal{Q}^{\pm}(\omega,\vecq)
= \mathcal{A}^{\pm}(\omega,\vecq)\mathcal{C}(\omega,\vecq)
   -\mathcal{B}(\omega,\vecq)\mathcal{D}^{\pm}(\omega,\vecq),\\
\tilde{\mathcal{Q}}^{\pm}(\omega,\vecq) 
= \tilde{\mathcal{A}}^{\pm}(\omega,\vecq)\mathcal{C}(\omega,\vecq)
   -\tilde{\mathcal{B}}(\omega,\vecq)\mathcal{D}^+(\omega,\vecq).
\eea
The computations of the phase-space integrals in Eqs.~
(\ref{eq:loopA}) to (\ref{eq:loopD}) can be
simplified~\cite{Leggett:1966zz}, because each loop can be written as a
product of some thermal function and a pre-factor that depends only on
the quantities $\omega$ and $\vecq\cdot \vecv_F$. To carry out the
phase-space integrations we first change the integration measure:
\be\label{intmeasure} 
\int\!\frac{d^3p}{(2\pi)^3} \simeq\nu\int
\frac{d\Omega}{4\pi} \int\limits_{-\infty}^{\infty}\!d\xi_p,
\ee
 where we used the fact that at low temperatures the lower integration limit
$-\mu/T \simeq -\infty$.  For the sake of completeness we list the
resulting expressions for the loops~\cite{Migdal:1967,Leggett:1966zz}
\bea
\label{eq:LoopA}
\mathcal{A}^{\pm}
&=&\nu\int\!\frac{d{{\Omega}}}{4\pi}\,
   \Biggl\{-\frac{1\pm\hat{\mathcal{P}}}{2}\,\mathcal{G}(\vecv,\omega,\vecq)
   \nonumber\\
  &&
  +\frac{\vecq\vecv}{\omega-\vecq\vecv}
   \,\Big[\mathcal{G}(\vecv,\vecq\vecv,\vecq)-\mathcal{G}(\vecv,\omega,\vecq)\Big]\Biggr\},
\\
\label{eq:LoopB}
\mathcal{B}
&=&-\nu\int\!\frac{d{{\Omega}}}{4\pi}\,
   \frac{\omega+\vecq\vecv}{2\Delta}\,\mathcal{G}(\vecv,\omega,\vecq),\\
\label{eq:LoopC}
\mathcal{C}
&=&\nu\int\!\frac{d{{\Omega}}}{4\pi}
   \frac{\omega^2-(\vecq\vecv)^2}{4\Delta^2}\,\mathcal{G}(\vecv,\omega,\vecq),\\
\label{eq:LoopD}
\mathcal{D}^\pm
&=&\nu\int\!\frac{d{{\Omega}}}{4\pi}\,
   \left[\frac{\omega+\vecq\vecv}{4\Delta}
        +\frac{\omega-\vecq\vecv}{4\Delta}\hat{\mathcal{P}}\right]\,
   \mathcal{G}(\vecv,\omega,\vecq),\nonumber\\
\eea
where the thermal function is given by 
\bea
\label{eq:Gthermal}
\mathcal{G}(\vecv,\omega,\vecq)
&=&
  \Delta^2\int\limits_{-\infty}^{+\infty}\!d\xi_p
\nonumber\\
&&\hspace{-1cm}\times
  \Biggl[\frac{\epsilon_+-\epsilon_-}{\epsilon_+\epsilon_-}
\frac{f(\epsilon_-)-f(\epsilon_+)}
{\omega^2-(\epsilon_+-\epsilon_-)^2+i\eta
  }\nonumber\\
&&\hspace{-1cm}-\frac{\epsilon_++\epsilon_-}{\epsilon_+\epsilon_-}
  \frac{1-f(\epsilon_-)-f(\epsilon_+)}{\omega^2-(\epsilon_++\epsilon_-)^2+i\eta}
  \Biggr],
\eea
where $f(x) = \{\exp[(x-\mu)/T]+1\}^{-1}$ is the fermionic
distribution function.
In the following we focus on the pair-breaking part of 
Eq. (\ref{eq:Gthermal})  given by
\bea
\mathcal{G}^{\pb}(\vecv,\omega,\vecq)
&=& -\Delta^2\int\limits_{-\infty}^{+\infty}\!d\xi_p
  \frac{(\epsilon_++\epsilon_-)}{\epsilon_+\epsilon_-}
\nonumber\\
&\times&  \frac{1-f(\epsilon_-)-f(\epsilon_+)}{\omega^2
-(\epsilon_++\epsilon_-)^2+i\eta},
\eea
which is the dominant part of the response in the low-temperature
domain.

\section{Thermal function}
\label{AppendixB}
\subsection{Analytical result}
Here we determine the real and imaginary parts of the zeroth
order coefficient in the expansion of the thermal function. The
first step is to use the generalized Dirac identity 
\bea
\int\!\frac{dz\,f(z)}{(z-\zeta+i\eta)^{n+1}}
&=&\dashint\!\frac{dz\,f(z)}{(z-\zeta)^{n+1}}\nonumber\\
&&\hspace{-2cm}-i\pi\,\frac{(-1)^n}{n!}\int\!dz\,f(z)\,\frac{\partial^n}{\partial z^n}\delta(z-\zeta)
\eea
and to decompose the complex function at hand into real and imaginary
parts. The imaginary part can be integrated analytically using the
partial integration in the formula 
\be
\int\limits_\alpha^\beta\!dz\,f(z)\,\delta^{(n)}(z-\zeta)
=(-1)^n\left.\frac{\partial^n f(z)}{\partial z^n}\right|_{z=\zeta}\,
\quad \forall \zeta\in[\alpha,\beta].
\ee
Once the imaginary part is calculated, the real part can  be
obtained via the Kramers-Kronig relation 
\be
{\rm Re}\,\phi(\omega)
=\frac{1}{\pi}\dashint\limits_{-\infty}^{+\infty}
 \frac{d\omega'\,{\rm Im}\phi(\omega')}{\omega'-\omega},
\ee
provided that 
the imaginary part decays faster than $1/\omega$ for large $\omega$.
The application of this procedure to the thermal function to leading 
order gives
\be
\mathcal{G}_0^{\rm pb} 
= -\Delta^2
  \int\limits_{-\infty}^{\infty}\!d\xi_p\,
  \frac{2\tanh\left(\frac{\epsilon_p}{2T}\right)}
       {\epsilon_p(\omega^2-4\epsilon_p^2+i\delta)}.
\ee
Next we compute the imaginary part and obtain 
\bea
\label{eq:ImG}
{\rm Im}\,\mathcal{G}_0^{\rm pb}
&=&\pi\Delta^2\,\int\limits_{-\infty}^{\infty}\!d\xi_p\,
   \frac{2\tanh\left(\frac{\epsilon_p}{2T}\right)}{\epsilon_p}
   \delta(\omega^2-4\epsilon_p^2)\nonumber\\
&=&\frac{2\pi\Delta^2}{|\omega|}
   \frac{\tanh\left(\frac{\omega}{4T}\right)}{\sqrt{\omega^2-4\Delta^2}}
   \,\theta\left(\frac{\omega}{2}-\Delta\right).
\eea
Note the threshold behavior enforced by the Heavyside function: Energy
transfer is possible only for frequencies larger than the
pair-breaking threshold $2\Delta$. Furthermore, the thermal function
at this order is independent of the momentum transfer; this implies
that the momentum-transfer dependence of the response functions is
determined by the pre-factors of the loop functions 
[c.f. Eqs.~(\ref{eq:LoopA}) to (\ref{eq:LoopD})].

\subsection{Numerical calculation of the thermal function}
In this section we  focus on the numerical calculation of the 
angle average of the thermal function, which is given by
\bea
{\cal G}^{\rm pb}(\vecv,\omega,\vecq)
&=&-\Delta^2\int\!\frac{d^3p}{(2\pi)^3}
\frac{(\epsilon_++\epsilon_-)}{\epsilon_+\epsilon_-}\nonumber\\
&\times&
\frac{(1-f(\epsilon_-)-f(\epsilon_+))}{(\omega^2-(\epsilon_++\epsilon_-)^2+i\eta)}\nonumber\\
&=&-2\Delta^2
   \int\limits_0^{2\pi}\!\frac{d\phi}{2\pi}
   \int\limits_{-1}^{+1}\!\frac{dx}{2}
   \int\limits_{\Delta}^{\infty}\!\frac{d\epsilon_p\,\epsilon_p}{\sqrt{\epsilon_p^2-\Delta^2}}
   \frac{(\ep_++\ep_-)}{\ep_+\ep_-}\nonumber\\
&\times&\frac{\left(1-f\left(\ep_-\right) -f\left(\ep_+\right)\right)}
        {\left(\omega^2-\left(\ep_++\ep_-\right)^2+i\eta\right)}.
\eea
The factor of 2 in the second relation arises from the fact that the
integrand is  an even function of $\xi_p$ and the integration can be
restricted to the positive values of the argument.  We have also
replaced the integration over the unpaired spectrum by the integration
over the paired spectrum by means of the relation $\xi_p d\xi_p =
\ep_p d\ep_p $. The integral over the azimuthal angle is trivial, since
the integrand is independent of $\phi$.  Further, after using the Dirac
identity we obtain
\bea
{\rm Im}\,{\cal G}^{{\rm pb}}(\vecv,\omega,\vecq)
&&=2\pi\Delta^2
  \int\limits_{-1}^{+1}\!\frac{dx}{2}
  \int\limits_{\Delta}^{\infty}\!\frac{d\epsilon_p\,\epsilon_p}{\sqrt{\epsilon_p^2-\Delta^2}}
  \frac{\left(\ep_++\ep_-\right)}
       {\ep_+\ep_-}\nonumber\\
&& \hspace{-2cm}\times
\left(1-f\left(\ep_+\right) -f\left(\ep_-\right)\right)
      \delta \left(\omega^2-\left(\ep_++\ep_-\right)^2\right).
\eea
One of the integrations can be carried out with the help of the
$\delta$ function. One finds 
\bea
{\rm Im}{\cal G}^{\rm pb}(\vecv,\omega,\vecq)
&=&\frac{\pi\Delta^2}{2}
  \int\limits_{\Delta}^{\infty}\!
  \frac{d\epsilon_p\,\epsilon_p}{\sqrt{\epsilon_p^2-\Delta^2}}
  \nonumber\\
&&\hspace{-3.6cm}\times \sum_{x_1,x_2} \frac{\left(\ep_++\ep_-\right)}
       {\ep_+\ep_-}
  \frac{\left[1-f\left(\ep_+\right) -f\left(\ep_-\right)\right]}
       {\vert(\ep_+ + \ep_{-})'\vert_{x=x_{1,2}}\vert}
  \theta\left(1-\vert x_{1,2}\vert \right).
\eea
where $x_{1,2}$ are the solutions of the equation
$\omega^2-\left[\ep_+(x)+\ep_-(x)\right]^2 = 0$ and the prime denotes a
derivative with respect to $x$; its explicit form is given 
elsewhere~\cite{Keller_Thesis}. Once the imaginary part is computed, 
the real part follows from the Kramers-Kronig relation. This completes 
our numerical procedure for computing the response functions. Each 
of the loops (\ref{eq:LoopA}) to (\ref{eq:LoopD}) can be computed by 
multiplying the numerical result for the thermal function by the 
appropriate pre-factor.

\section{Comparison to Leggett's formalism }
\label{AppendixC}
Here we compare the response functions derived above with the results
obtained in the Leggett formalism~\cite{Leggett:1966zz,Leinson:2009mq}
and establish the correspondence between the two.  The full
(effective) normal vertices in the Leggett's formalism
~\cite{Leggett:1966zz} are defined as symmetrical and anti-symmterical
combinations of the particle $\tau$ and hole $\tau^{h}$ vertices \be
\label{eq:tau_eff}
 \tau^{+}
=\frac{1}{2}\big(\tau+\tau^{h}\big)\,,\quad 
 \tau^{-}
=\frac{1}{2}\big(\tau-\tau^{h}\big).
\ee
If $\tau = \pm \tau^{h}$, \ie, the vertices have  odd or even parity
under transformations which convert particles into holes, then one 
of the linear combinations (\ref{eq:tau_eff}) vanishes. The bare
vertices are defined analogously 
\be
 \xi^{+}=\frac{1}{2}\big(\xi+\xi^{h}\big)\,,\quad 
 \xi^{-}=\frac{1}{2}\big(\xi-\xi^{h}\big)\,.
\ee
The anomalous vertex is denoted by $\tilde{\tau}$. With these
definitions the integral equations for the full vertices are 
\bea
\label{eq:L1}
&&\Big[1-\int\frac{d\Omega'}{4\pi}\,V_{\pp}^{D/S}(\vecv\vecv')\,A_0\nonumber\\
& &      +\int\frac{d\Omega'}{4\pi}\,V_{\pp}^{D/S}(\vecv\vecv')\,\frac{\omega^2-(\vecq\vecv')^2}{2\Delta^2}\,\lambda(\vecv')\Big]\,\tilde{\tau}
 \nonumber\\
 &
 +&\int\frac{d\Omega'}{4\pi}\,V_{\pp}^{D/S}(\vecv\vecv')\,\frac{\vecq\vecv'}{\Delta}\,\lambda(\vecv')\,\tau^{-}
\nonumber\\
& -&\int\frac{d\Omega'}{4\pi}\,V_{\pp}^{D/S}(\vecv\vecv')\,\frac{\omega}{\Delta}\,\lambda(\vecv')\,\tau^{+}
 =0\,,\\
\label{eq:L2}
&&\int\frac{d\Omega'}{4\pi}\,V_{\ph}^{D/S}(\vecv\vecv')\,\frac{\vecq\vecv'}{\Delta}\,\lambda(\vecv')\,\tilde{\tau}
 \nonumber\\
 &
 +&\Big[1-\int\frac{d\Omega'}{4\pi}\,V_{\ph}^{D/S}(\vecv\vecv')\,\kappa(\vecv')\Big]\,\tau^{-}
\nonumber\\
& +&\int\frac{d\Omega'}{4\pi}\,V_{\ph}^{D/S}(\vecv\vecv')\,\frac{\omega}{\vecq\vecv'}\,\kappa(\vecv')\,\tau^{+}
 =\xi^{-}\,,\\
\label{eq:L3}
&-&\int\frac{d\Omega'}{4\pi}\,V_{\ph}^{D/S}(\vecv\vecv')\,\frac{\omega}{\Delta}\,\lambda(\vecv')\,\tilde{\tau}
 \nonumber\\
 &
 +&\int\frac{d\Omega'}{4\pi}\,V_{\ph}^{D/S}(\vecv\vecv')\,\frac{\omega}{\vecq\vecv'}\,\kappa(\vecv')\,\tau^{-}
\nonumber\\
 &
 +&\Big[1-\int\frac{d\Omega'}{4\pi}\,V_{\ph}^{D/S}(\vecv\vecv')\,(\kappa(\vecv')-2\lambda(\vecv')\Big]\,\tau^{+}
 =\xi^{+},\nonumber\\
 \eea
where 
\bea
\label{eq:kappa}
 \kappa(\vecv')
&=&\frac{1}{2}\Big[G^{h}_{+}G^{h}_{-}+G_{+}G_{-}\Big]+F_{+}F_{-}\,,\\ 
\label{eq:lambda}
 \lambda(\vecv')
&=&F_{+}F_{-}\,,\\
\label{eq:A0}
A_0 &=& -\lim_{w\rightarrow0,\vecq\rightarrow0}\Big[GG^{h}+FF\Big]\,,
 \eea
where, as before, the wave-function renormalization is set to
unity. Keeping only the lowest order term in the expansion of the
particle-particle interaction in Eq. (\ref{eq:L1}), one finds $
1-V_{\pp}^{0}A_0=0$, \ie, the first two terms in that equation mutually
cancel. (Note  the different sign convention for $V_{\pp}^0$ in the
main body of the paper.)

We proceed now to solve these equations for the vertices in some cases
of interest. For that purpose define the following integrals: \bea
\alpha
&=&\int\frac{d\Omega'}{4\pi}\,V_{\ph}^{D/S}(\vecv\vecv')\,\lambda(\vecv')\,,\\
\eta
&=&\int\frac{d\Omega'}{4\pi}\,V_{\ph}^{D/S}(\vecv\vecv')\,\kappa(\vecv')\,,\\
\gamma
&=&\int\frac{d\Omega'}{4\pi}\,V_{\ph}^{D/S}(\vecv\vecv')\,\cos^2\theta\,\lambda(\vecv')\,,\\
\beta
&=&\int\frac{d\Omega'}{4\pi}\,V_{\ph}^{D/S}(\vecv\vecv')\,\cos^2\theta\,\kappa(\vecv')\,,\\
\psi
&=&\int\frac{d\Omega'}{4\pi}\,V_{\ph}^{D/S}(\vecv\vecv')\,\cos\theta\,\lambda(\vecv')\,,\\
\phi
&=&\int\frac{d\Omega'}{4\pi}\,V_{\ph}^{D/S}(\vecv\vecv')\,\cos^{-1}\theta\,\lambda(\vecv')\,.
\eea 
In the following, we keep (as in the main body of this paper) the
leading order Landau parameter in the particle-hole interaction
amplitude, \ie, $V_{\ph}^{D/S}(\vecv\vecv')=v_{\ph}$. Because 
 $\kappa$ and $\lambda$ are even functions of
$\cos\theta$, in this approximation the functions $\phi$ and $\psi$
vanish. Following Leggett~\cite{Leggett:1966zz}, we will use below the
abbreviations \be s=\frac{\omega}{qV_F}\quad u=\frac{qV_F}{\Delta}.
\ee

The longitudinal component in the vector channel,  is obtained when
$\xi^{+}=1$ and $\xi^{-}=0$. Equations (\ref{eq:L1}) to (\ref{eq:L3}) are then
written as
\bea
\left( \begin{array}{ccc}
  s^2\,\alpha-\gamma    
 &2\,u\,\psi
 &-2\,su\,\alpha
 \\ 
  v_{\ph}\,u\,\psi
 &1-v_{\ph}\,\eta
 &v_{\ph}\,s\,\phi
 \\
  -v_{\ph}\,su\,\alpha
 &v_{\ph}\,s\,\phi
 &1-v_{\ph}\,(\eta-2\alpha)
 \end{array}\right)\left(
 \begin{array}{ccc}
 \tilde{\tau} 
 \\
 \tau^{-}
 \\
 \tau^{+}
 \end{array}\right)
 =
\left( \begin{array}{ccc}
 0 
 \\
 0
 \\
 1
 \end{array}\right)\,,
\nonumber
\eea
As stated above $\phi = \psi = 0$ at leading order in the
particle-hole interaction.  The solution of this matrix equation 
is given by
\bea
 \tilde{\tau}
&=&\frac{2\Delta}{\omega}\,
  \frac{s^2\alpha}{(s^2\alpha-\gamma)\big[1-v_{\ph}\,Q_L\big]}\,,\\
 \tau^{+}
&=& \frac{1}{1-v_{\ph}\,Q_L}\,,
\eea
and $\tau^{-}=0$, where
\begin{equation}
\label{eq:QL}
 Q_L\equiv \eta+\frac{2\alpha\gamma}{(s^2\alpha-\gamma)}\,, 
\end{equation}
in agreement with  Eqs.~(52), (53) and (54) of
Ref.~\cite{Leinson:2009mq}  taken in the case of $V_{\ph}^{1}= 0$.

The longitudinal projection of the vector current polarization tensor
is given by the expression [cf.~\cite{Leggett:1966zz}, Eq. (23a)]
\begin{equation}
 \Pi_{V,L}
=\int\frac{d\Omega}{4\pi}\,
 \xi^{+}\,
 \left[\frac{\omega}{\Delta}\lambda\,\tilde{\tau}
      +\frac{\omega}{qV_F}\lambda\,\tau^{-}
      +(\kappa-2\lambda)\,\tau^{+}\right], 
\end{equation}
which after the substitution of the vertices becomes 
\begin{equation}
 \Pi_{V,L}
=
 \,\left[\eta+\frac{2\alpha\gamma}{(s^2\alpha-\gamma)}\right]
 \,\frac{1}{1-v_{\ph}\,Q_L}
=\frac{Q_L}{1-v_{\ph}\,Q_L}\,.
\end{equation}
By matching this equation to our result given by Eq. (\ref{eq:VL}) we
find 
\begin{equation}
\label{eq:match}
 Q_{L}
 =\mathcal{A}^{+}-\frac{\mathcal{D}^{+}}{\mathcal{C}}\mathcal{B}
 =\frac{\mathcal{Q}^{+}}{\mathcal{C}}\,.
\end{equation}
The latter equality is straightforward to prove by noting that
Eqs. (\ref{eq:lambda}) and (\ref{eq:kappa}) can be written in terms of
the thermal function (\ref{eq:Gthermal}) as 
\bea 
\label{eq:lmatch}
\lambda
&=&\frac{\nu}{2}\,\mathcal{G}(\vecv,\omega,\vecq)\,,\\
\label{eq:kmatch}
\kappa &=&\nu\Biggl\{\frac{(\vecq\vecv)^2}{\omega^2-(\vecq\vecv)^2}
\,\Big[\mathcal{G}(\vecv,\vecq\vecv,\vecq)-\mathcal{G}(\vecv,\omega,\vecq)\Big]\Biggr\}\,.\nonumber\\
\eea 
We conclude that our result for the longitudinal vector current
response function agrees with those given in
Refs.~\cite{Leggett:1966zz,Leinson:2009mq}.  In particular, we have
verified that the limiting cases of (i) $v_{\ph} = 0$, (ii) $\omega\ll
\Delta$, $qv_F\ll \Delta$ for non-zero $T$, and (iii) same as in (ii),
but for $T=0$, we recover the results of Ref.~\cite{Leinson:2009mq} by
using the matching condition (\ref{eq:match}).  However, the
perturbative result for the imaginary part of the longitudinal vector current
response function in Ref.~\cite{Leinson:2009mq} [second term in
Eq. (82)] differs from our result, given by Eqs.~(\ref{eq:pi_density})
and (\ref{eq:ImG}) by a factor of 1/8. (Note that the author
Ref.~\cite{Leinson:2009mq} used a density of state which is by a
factor of 2 larger than ours). We have verified that one recovers our
result by starting from the exact expression (\ref{eq:QL}) for $Q_L$
and expanding the functions $\alpha$, $\eta$ and $\gamma$ in small
$s^{-1}$. In Ref. ~\cite{Leinson:2009mq} the exact expression is first
approximated by $Q_L \simeq \eta + 2 s^{-2}\gamma$ after which the
expansions for $\eta$ and $\gamma$ are substituted. The first step
is the source of the discrepancy; we have verified that if the
expansions of the functions  $\alpha$, $\eta$ and $\gamma$ are
directly substituted in the exact expression  (\ref{eq:QL}) for $Q_L$,
then one recovers our result, which is also in agreement with the one
quoted earlier by the authors of Refs.~\cite{Kolomeitsev:2010hr,Kolomeitsev:2008mc}.

In the case of the current response the bare vertices are given by
$\xivec=\vecv_\bot$ and $\xivec^h=-\vecv_\bot$, \ie, \be
\xivec^{+}=0\,,\quad \xivec^{-}=\vecv_\bot \ee where $\vecv_\bot$
denotes the transverse to the momentum transfer projection of the
quasiparticle velocity. If the momentum transfer is along the $z$
axis, then
$\vecv_\bot=v_F(\sin\theta\cos\phi,\sin\theta\sin\phi,\cos\theta)\,.$
Keeping only the leading order Landau parameter in the particle-hole
channel one finds \bea \tilde{\tauvec}^{(i)}=0\,,\quad
\tauvec^{+(i)}=0\,,\quad \tauvec^{-(i)}=\vecv_\bot^{(i)}\,.  \eea The
anomalous vertex vanishes identically (the contributions to the vertex
in the direction of the momentum transfer are neglected here).  After
substituting the vertices into the expression for the transverse part
of the polarization tensor (Eq. (37) in Ref.~\cite{Leinson:2009mq}) we
find \be\label{eq:Ltrans}
\Pi_{V,T}=\frac{v_F}{2}\int\limits_\Omega\,\tau^{-}\kappa\,\big(1-\cos^2\theta\big)
=\frac{v_F^2}{2}\big(\eta-\beta\big)\,, \ee which coincides with
Eq. (86) of Ref.~\cite{Leinson:2009mq} when $v_{\ph}^{1}=0$.  The
transverse vector response function is given according to
Eq.~(\ref{eq:res_VT}) above.  After substituting the explicit
expression for the loop function ${\cal A}^-$ from
Eq. (\ref{eq:LoopA}) we use the relation (\ref{eq:kmatch}) to recover
Eq. (\ref{eq:Ltrans}), \ie, the transverse vector polarization tensors
are the same in both approaches. However, the perturbative expansions
of these transverse polarization tensors differ, by a factor $O(1)$,
c.f. Eq. (88) in Ref.~\cite{Leinson:2009mq} and
Eqs.~(\ref{eq:pi_current}) and (\ref{eq:ImG}) above.

The longitudinal and transverse axial-vector current polarization
tensors of Ref.~\cite{Leinson:2009mq}  can be matched to our results 
as in the case of vector current response functions, therefore we do
not repeat the arguments above.


\begin{thebibliography}{99}
\bibitem{Baym:1978jf} 
 G.~Baym and C.~Pethick,
 Ann.\ Rev.\ Astron.\ Astrophys.\  {\bf 17}, 415 (1979).

\bibitem{Pethick:1995di} 
 C.~J.~Pethick and D.~G.~Ravenhall,
 Ann.\ Rev.\ Nucl.\ Part.\ Sci.\  {\bf 45}, 429 (1995).


\bibitem{Lombardo:2000ec} 
  U.~Lombardo and H.~J.~Schulze,
  Lect.\ Notes Phys.\  {\bf 578}, 30 (2001).

\bibitem{Dean:2002zx} 
  D.~J.~Dean and M.~Hjorth-Jensen,
  Rev.\ Mod.\ Phys.\  {\bf 75}, 607 (2003).

\bibitem{Sedrakian:2006xm}
 A.~Sedrakian and J.~W.~Clark,
in {\it  ``Pairing in Fermionic Systems: Basic Concepts and Modern
applications''}, 
World Scientific,  Singapore, 2006, p. 135.

\bibitem{Pwave1}
M. Baldo, O. Elgar\o{}y, L. Engvik, M. Hjorth-Jensen, H.-J. Schulze, 
Phys. Rev. C {\bf 58}, 1921  (1998).

\bibitem{Pwave2}
 M. V. Zverev, J. W. Clark, V. A. Khodel, Nucl. Phys. A {\bf 720}, 20
 (2003).

\bibitem{Pwave3}
  V. A. Khodel, J. W. Clark, M. V. Zverev,
 Phys. Rev. Lett. {\bf 87}, 031103 (2001).


\bibitem{Alm:1996zz} 
  T.~Alm, G.~R\"opke, A.~Sedrakian and F.~Weber,
  Nucl.\ Phys.\ A {\bf 604}, 491 (1996).




\bibitem{Olsson:2002yu} 
  E.~Olsson and C.~J.~Pethick,
  Phys.\ Rev.\ C {\bf 66}, 065803 (2002).

\bibitem{Olsson:2004ea}
 E.~Olsson, P.~Haensel and C.~J.~Pethick,
 Phys.\ Rev.\  C {\bf 70}, 025804 (2004).



\bibitem{Lykasov:2005xh}
 G.~I.~Lykasov, E.~Olsson and C.~J.~Pethick,
 Phys.\ Rev.\  C {\bf 72}, 025805 (2005).


\bibitem{Lykasov:2008yz} 
  G.~I.~Lykasov, C.~J.~Pethick and A.~Schwenk,
  Phys.\ Rev.\ C {\bf 78}, 045803 (2008).


\bibitem{Pethick:2009gj} 
  C.~J.~Pethick and A.~Schwenk,
  Phys.\ Rev.\ C {\bf 80}, 055805 (2009).



\bibitem{Margueron:2005nc}
J.~Margueron, N.~V. Giai and J.~Navarro,
Phys.\ Rev.\  C {\bf 72}, 034311 (2005).

\bibitem{Bozek:2004ct} 
P.~Bozek, J.~Margueron and H.~M\"uther,
Annals Phys.\ (NY)\ {\bf 318}, 245 (2005).

\bibitem{Negele:1988vy}
J.~W.~Negele and H.~Orland,
{\it ``Quantum Many Particle Systems,''}
{(Addison-Wesley, New York,  1988).}

\bibitem{Friman:1978zq} 
 B.~L.~Friman and O.~V.~Maxwell,
 Astrophys.\ J.\  {\bf 232}, 541 (1979).

\bibitem{Sedrakian:2000kc} 
 A.~Sedrakian and A.~E.~L.~Dieperink,
 Phys.\ Rev.\ D {\bf 62}, 083002 (2000).

\bibitem{Hanhart:2000ae} 
 C.~Hanhart, D.~R.~Phillips and S.~Reddy,
 Phys.\ Lett.\ B {\bf 499}, 9 (2001).



\bibitem{Timmermans:2002hc} 
R.~G.~E.~Timmermans, A.~Y.~Korchin, E.~N.~E.~van Dalen and A.~E.~L.~Dieperink,
Phys.\ Rev.\ C {\bf 65}, 064007 (2002).



\bibitem{Sedrakian:2010xe} 
  A.~Sedrakian and J.~Keller,
  Phys.\ Rev.\ C {\bf 81}, 045806 (2010)


\bibitem{AG} A. A. Abrikosov, L. P. Gorkov, 
Sov. Phys. JETP {\bf 8}, 1090 (1959);  Sov. Phys. JETP {\bf 9}, 220 (1959).

\bibitem{Abrikosov:1962} A. A. Abrikosov, L. P. Gorkov, and I. E. Dzyaloshinski,
{\it  Methods of quantum field theory in statistical physics}, (Dover, New
York, 1975).

\bibitem{Bogolyubov} N. Bogolyubov, Nuovo Cimento {\bf 7}, 794 (1958).

\bibitem{Anderson}  P. W. Anderson, Phys. Rev. {\bf 110}, 827 (1958).

\bibitem{Migdal:1967} 
            A. I. Larkin and A. B. Migdal, Sov. Phys. JETP {\bf 17}, 1146 (1963);
            A. B. Migdal, {\it Theory of Finite Fermi
           Systems and applications
           to Atomic Nuclei} ~(Interscience, London, 1967).

\bibitem{Leggett:1966zz}
A.~J.~Leggett,
Phys.\ Rev.\  {\bf 147}, 119 (1966).

\bibitem{Sedrakian:2006mq}
A.~Sedrakian,
Prog.\ Part.\ Nucl.\ Phys.\  {\bf 58}, 168-246 (2007).


\bibitem{Kundu:2004mz} 
J.~Kundu and S.~Reddy,
Phys.\ Rev.\ C {\bf 70}, 055803 (2004).

\bibitem{Leinson:2006gf} 
  L.~B.~Leinson and A.~Perez,
  Phys.\ Lett.\ B {\bf 638}, 114 (2006).

\bibitem{Steiner:2008qz}
A.~W.~Steiner, S.~Reddy,
Phys.\ Rev.\  {\bf C79}, 015802 (2009).


\bibitem{Sedrakian:2006ys}
A.~Sedrakian, H.~M\"uther, P.~Schuck,
Phys.\ Rev.\  {\bf C76}, 055805 (2007).




\bibitem{Kolomeitsev:2008mc}
E.~E.~Kolomeitsev, D.~N.~Voskresensky,
Phys.\ Rev.\  {\bf C77}, 065808 (2008).


\bibitem{Leinson:2009mq} 
  L.~B.~Leinson,
  Phys.\ Rev.\ C {\bf 79}, 045502 (2009).

\bibitem{Kolomeitsev:2010hr}
 E.~E.~Kolomeitsev and D.~N.~Voskresensky,
  Phys.\ Rev.\  C {\bf 81}, 065801 (2010).  



\bibitem{Sedrakian:2012ha} 
 A.~Sedrakian,
Phys.\ Rev.\ C {\bf 86}, 025803 (2012).


\bibitem{Pethick:2010zf} 
  C.~J.~Pethick, N.~Chamel and S.~Reddy,
  Prog.\ Theor.\ Phys.\ Suppl.\  {\bf 186}, 9 (2010).

\bibitem{Cirigliano:2011tj} 
 V.~Cirigliano, S.~Reddy and R.~Sharma,
 Phys.\ Rev.\ C {\bf 84}, 045809 (2011).

\bibitem{Sedrakian:1996} 
A. Sedrakian,  Astrophys. and Space Sci. {\bf 236}, 267 (1996).

\bibitem{DiGallo:2011cr} 
 L.~Di Gallo, M.~Oertel and M.~Urban,
 Phys.\ Rev.\ C {\bf 84}, 045801 (2011).

\bibitem{Carter:2004zr} 
  B.~Carter, N.~Chamel and P.~Haensel,
  Nucl.\ Phys.\ A {\bf 759}, 441 (2005).


\bibitem{Carter:2006nd} 
  B.~Carter and E.~Chachoua,
  Int.\ J.\ Mod.\ Phys.\ D {\bf 15}, 1329 (2006).

\bibitem{Carter:2006mv} 
  B.~Carter and L.~Samuelsson,
  Class.\ Quant.\ Grav.\  {\bf 23}, 5367 (2006).





\bibitem{Baldo:2008pb}
 M.~Baldo and C.~Ducoin,
 Phys.\ Rev.\  C {\bf 79}, 035801 (2009).

\bibitem{Baldo:2011nc}
M.~Baldo and C.~Ducoin,
Phys.\ Atom.\ Nucl.\  {\bf 74}, 1508 (2011).

\bibitem{Baldo:2011iz}
M.~Baldo and C.~Ducoin,
Phys.\ Rev.\  C {\bf 84}, 035806 (2011).

\bibitem{Flowers:1976ux}
E.~Flowers, M.~Ruderman, P.~Sutherland,
Astrophys.\ J.\  {\bf 205}, 541 (1976).

\bibitem{Yakovlev:1998wr} 
  D.~G.~Yakovlev, A.~D.~Kaminker and K.~P.~Levenfish,
  Astron.\ Astrophys.\  {\bf 343}, 650 (1999).

\bibitem{Kaminker:1999ez}
A.~D.~Kaminker, P.~Haensel, D.~G.~Yakovlev,
Astron.\ Astrophys.\  {\bf 345}, L14-L16 (1999).

\bibitem{Vaks}
V.~G.~Vaks, V.~M.~Galitski, A.~I.~Larkin, Sov. Phys. JETP {\bf 14} 1177 (1962).

\bibitem{Kolomeitsev:2011wz} 
  E.~E.~Kolomeitsev and D.~N.~Voskresensky,
  Phys.\ Rev.\ C {\bf 84}, 068801 (2011).


\bibitem{Keller_Thesis} 
J. Keller, Ph. D. Thesis, Frankfurt am Main, 2013.

\bibitem{Voskresensky:1987hm}
D.~N.~Voskresensky, A.~V.~Senatorov, Yad. Fiz. {\bf 45},
411 (1987)
[Sov.\ J.\ Nucl.\ Phys.\  {\bf 45}, 657  (1987)].


\bibitem{Ashcroft}
N.~W.~Ashcroft and N.~D.~Mermin,
 {\it Solid State Physics} (Saunders College, Philadelphia, 
1976).


\end{thebibliography}
\end{document}